\documentclass[aps,pra,twocolumn,showpacs,superscriptaddress]{revtex4-1}

\usepackage{ifpdf}
\usepackage{amsmath} 
\usepackage{amssymb}
\usepackage{amsfonts}
\usepackage{braket}
\usepackage{graphicx}
\usepackage{color}
\usepackage[colorlinks=true,linkcolor=blue,citecolor=blue,breaklinks]{hyperref}

\newcommand{\Hc}{{\mathrm{H.c.}}}
\newcommand{\up}{{\uparrow}}
\newcommand{\dn}{{\downarrow}}
\def\etal{~\textit{et.~al}}

\begin{document}

\title{Quasiparticle explanation of ``weak thermalization" regime under quench in a non-integrable quantum spin chain}

\author{Cheng-Ju Lin}
\affiliation{Department of Physics and Institute for Quantum Information and Matter, California Institute of Technology, Pasadena, CA 91125, USA}
\author{Olexei I. Motrunich}
\affiliation{Department of Physics and Institute for Quantum Information and Matter, California Institute of Technology, Pasadena, CA 91125, USA}

\date{\today}

\begin{abstract}
The eigenstate thermalization hypothesis provides one picture of thermalization in a quantum system by looking at individual eigenstates.
However, it is also important to consider how local observables reach equilibrium values dynamically.
Quench protocol is one of the settings to study such questions.
A recent numerical study [Ba\~{n}uls, Cirac, and Hastings, Phys.~Rev.~Lett.~\textbf{106}, 050405 (2011)] of a nonintegrable quantum Ising model with longitudinal field under such a quench setting found different behaviors for different initial quantum states.
One particular case called the ``weak thermalization" regime showed apparently persistent oscillations of some observables.
Here we provide an explanation of such oscillations.
We note that the corresponding initial state has low energy density relative to the ground state of the model.
We then use perturbation theory near the ground state and identify the oscillation frequency as essentially a quasiparticle gap.
With this quasiparticle picture, we can then address the long-time behavior of the oscillations.
Upon making additional approximations which intuitively should only make thermalization weaker, we argue that the oscillations nevertheless decay in the long time limit.
As part of our arguments, we also consider a quench from a BEC to a hard-core boson model in one dimension.
We find that the expectation value of a single-boson creation operator oscillates but decays exponentially in time, while a pair-boson creation operator has oscillations with a $t^{-3/2}$ decay in time.
We also study dependence of the decay time on the density of bosons in the low-density regime and use this to estimate decay time for oscillations in the original spin model.
\end{abstract}


\maketitle

\section{Introduction}

Pioneered by Boltzmann, statistical mechanics has been hugely successful in describing our physical world with many degrees of freedom.
The key idea is that for a thermodynamically large system, we only need few parameters to describe the system and do not need to know details of the microscopic dynamics.
In classical physics, it is known that the chaotic dynamics in the phase space leads to ergodicity, hence the validity of statistical mechanics.
On the other hand, in quantum physics, it was not clear how to draw the connection between quantum mechanics and statistical mechanics.
Two decades ago, Srednicki \cite{Srednicki1994} and Deutch \cite{Deutsch1991} independently proposed a mechanism to illustrate how statistical mechanics can emerge in a quantum system.
This hypothesis, known as the eigenstate thermalization hypothesis (ETH), essentially proposes that for any ``thermalizing" Hamiltonian, each eigenstate at a finite energy density itself is a superposition with random phases in some local observable's eigenbasis.
Therefore, when one does any local measurement, the eigenstate itself can already produce what looks like an ensemble average, without considering time averages required in classical physics.
Many studies have been published to support this conjecture \cite{Kim2014,beugeling_finite-size_2014, Garrison2015,beugeling_global_2015, Mondaini2016} and its connection to microcanonical and canonical ensembles \cite{Rigol2008}.

While the ETH can provide an explanation or a criterion to determine if a system will thermalize or not at infinite time, how the system evolves into such an ``equilibrium" or even thermalizes as a function of time is another important and challenging question.
The increasing interest in non-equilibrium dynamics of quantum many-body systems is also stimulated by cold atom experiments \cite{Greiner2002, Kinoshita2006, Hofferberth2007, Trotzky2012, Cheneau2012, Gring2012}, where one can control the initial state and the dynamical Hamiltonian and study real-time dynamics.  
On the other hand, theoretical studies to-date rely heavily on numerical studies \cite{Rigol2007,Flesch2008a,beugeling_off-diagonal_2015,Collura2015,mazza_overlap_2016}. 
Analytical results are limited primarily to integrable lattice models\cite{Rossini2010, Calabrese2011, Calabrese2012, Calabrese2012a, Kormos2014, Piroli2016}, in which the equilibrium ensemble is believed to be described by a generalized Gibbs ensemble, and continuum field theory near the critical point~\cite{Delfino2014, Delfino2016}.
Models with weak integrability-breaking interactions have also been studied, where a ``prethermalization" stage was established at the intermediate time scale before the eventual thermalization~\cite{Berges2004,Maurizio2014,Bruno2015,Tim2016, Marcuzzi2016}.
Prethermalized states of Bose gases and Fermi gases close to Feshbach-resonance were considered in Refs. \cite{yin_postquench_2016,yin_quench_2016}.
We refer readers to Refs.~\cite{Eisert2015, Essler2016} for recent more comprehensive reviews.

Despite the widely held belief that a non-integrable model will generally thermalize, a recent numerical study by Ba\~{n}uls, Cirac, and Hastings \cite{Banuls2011} observed some unusual and interesting behaviors.
The authors used an infinite-matrix-product-state (infinite-MPS) technique \cite{Banuls2009} to study the following quench problem.
Starting from various initial product states, one measures local observables as a function of time evolving under a generic non-integrable quantum spin Hamiltonian
\begin{equation}
H = -J \sum_{j=1}^L \sigma_j^z \sigma_{j+1}^z - h \sum_{j=1}^L \sigma_j^z - g \sum_{j=1}^L \sigma_j^x ~.
\label{H}
\end{equation}
For an initial state where all spins are pointing in the $\hat{y}$ direction, $|Y\!+ \rangle$, the behavior is consistent with the conventional thermalization wisdom.
However, there are some initial states that display apparently different behaviors.
One type of behavior occurs for initial states with spins pointing close to the $\hat{z}$ direction, $|Z\!+ \rangle$, where observables show strong oscillations without damping for the entire time where the numerical simulation is reliable;
since the time-averaged observables apparently approach the thermal values, this behavior was called ``weak thermalization.'' 
A subsequent work \cite{Hastings2015} using an improved ``hybrid algorithm'' also found similar persistent oscillations starting from a different initial state $|X\!- \rangle$.
On the other hand, another type of behavior occurs for initial states close to $|X\!+ \rangle$, where a local observable $\sigma_j^x$ apparently does not thermalize, also upon time-averaging.

In this paper, we provide a simple quasiparticle explanation for the strong oscillation behavior observed in the ``weak thermalization'' case in Ref.~\cite{Banuls2011}.
We focus on the $|Z\!+ \rangle$ initial state and argue that it is actually close to the ground state of the above Hamiltonian, and the oscillation frequency can be essentially understood as the quasiparticle energy above this ground state.
This initial state has a finite energy density above the ground state and hence has a finite density of such quasiparticles.
However, when the quasiparticle density is small, the quasiparticles are effectively weakly interacting, and the oscillations in the observables can persist to long times.
Armed with this quasiparticle description of the origin of oscillations, we can then argue that the interactions among the quasiparticles will make the oscillations decay eventually.

The quasiparticle description developed here also leads us to consider the following quench problem, which is interesting on its own.
We argue that the quench problem starting from $|Z\!+ \rangle$ can be viewed approximately as a quench from a Bose-Einstein condensate (BEC) state evolving under a hard-core boson Hamiltonian.
The observable of interest can be viewed as a BEC order parameter, which exhibits the strong oscillation.

In fact, the above quench setting is essentially close to a quench from a magnetically ordered state in the quantum Ising chain \cite{Rossini2010, Calabrese2011, Calabrese2012, Calabrese2012a, Essler2016}, where it has been established that the magnetization order parameter decays exponentially in time.
The decoherence time of the order parameter was obtained analytically, and the mechanism for the decoherence can be understood as a destructive interference from Jordan-Wigner fermions with all momenta that are produced by the action of the order parameter field, which is non-local in terms of these fermions.

The BEC quench setting has been studied experimentally \cite{Greiner2002} and theoretically \cite{Kormos2014, Mazza2014}.
The previous studies focused on the evolution of a BEC state with a fixed number of particles, which is natural in experiments but also makes the evolution of the BEC order parameter more challenging to study. 
Indeed, to study $\langle b_j(t) \rangle$ at time $t$ in this setting, one needs to consider boson correlation $\langle b_j^\dagger(t) b_{j+\ell}(t) \rangle$ in the limit where the separation $\ell \rightarrow \infty $ first. 
More specifically, under hard-core boson Hamiltonian, the correlation function in the Jordan-Wigner fermion representation becomes an infinite-length string operator, which is a formidable calculation without Wick's theorem, as is the case for simple BEC states.

We perform essentially the above correlation function calculation upon using a further trick where we replace the simple product BEC initial state with a different state in the same phase but satisfying Wick's theorem for the Jordan-Wigner fermions.
Under these further choices of the initial state and the evolution Hamiltonian, we show that the BEC order parameter $\langle b_j(t) \rangle$ decays exponentially in time.
We also study how the decay rate depends on the density of particles in the initial state.
From the analogy with the quench in the quantum Ising model, we conjecture that similar expressions as in Refs.~\cite{Calabrese2012, Essler2016} for the decoherence time and the oscillation frequency will be applicable to our BEC quench setting.
We find that our numerical results are consistent with the conjectured expressions.
In particular, we find that the decay rate (i.e., inverse decoherence time) vanishes as $\rho^2 \ln(\frac{1}{\rho})$ at low densities $\rho$.

We remark that a recent work~\cite{Kormos2016} suggested a possible physics that could dramatically alter the conventional "light-cone" picture of the quench problem.
The Hamiltonian Eq.~(\ref{H}) is integrable for vanishing longitudinal field, $h = 0$, and thermalizes readily to the corresponding generalized Gibbs ensemble.
When the transverse field is below critical, $|g| < 1$, the propagating quasiparticles can be thought of as individual domain walls in the ferromagnetic order.
However, for non-zero longitudional field, $|h| > 0$, these domain walls are confined, which leads to a dramatic suppression of the light-cone propagation and the entanglement entropy growth observed in Ref.~\cite{Kormos2016}.
For small $h$, the true quasiparticles above the ground state can be thought of as "mesons," which are bound (confined) states of two domain walls. 
Reference~\cite{Kormos2016} calculated masses of stable such mesons for small $h$, and found that observables show apparently persistent oscillations with frequencies set by these masses.
Other authors, Refs.~\cite{Delfino2014, Delfino2016}, also proposed that oscillation frequencies in quantum quenches in near-critical one-dimensional (1D) systems are determined by quasiparticle masses.
In this respect, our quasiparticle explanation of oscillations in the weak thermalization regime is close in spirit to Refs.~\cite{Kormos2016, Delfino2014, Delfino2016}.
The microscopic details of the true quasiparticles are different in our regime with fairly large both $g$ and $h$ parameters, but this is a quantitative rather than qualitative difference with Ref.~\cite{Kormos2016}.
Our emphasis in this paper is more on the generic statement that the low-energy spectrum can be described in terms of particle-like excitations ("quasiparticles"), whose properties can be extracted, e.g., from ED studies, and that we should always think in terms of such quasiparticles when the initial state has low energy density over the ground state.
Our main development is the (approximate) picture of the initial state as a BEC of the quasiparticles and how this system eventually thermalizes (the "condensate" decays), which we argue implies that the physical observables cannot have persistent oscillations at long times.

The paper is organized as follows:
In Sec.~\ref{finitesize}, we compare our finite-size exact diagonalization (ED) results to the infinite-MPS results of Ref.~\cite{Banuls2011}.
The finite-size ED spectra enable us to identify the origin of the oscillation frequency, which we argue is essentially the energy of the quasiparticle.
In Sec.~\ref{perturbative}, we use Schrieffer-Wolff (SW) transformation to derive an effective Hamiltonian for the quasiparticles, which looks like a hard-core boson Hamiltonian with additional correlated hopping and interactions.
We further argue that in the SW-transformed picture, the initial state becomes a state with a non-zero BEC order parameter with low particle density, while the observables of interest will have the main components changing the particle number by one, which is the source of the oscillation frequency.
Components changing the particle number by more than one appear in higher-order.
Based on this identification, we further simplify the whole problem as a quench from a BEC state evolving under an integrable hard-core boson Hamiltonian with hopping only, which we study separately in Sec.~\ref{sec:quenchMatter}.
We consider two different initial states, both with nonzero BEC order parameter, in Secs.~\ref{productini}~and~\ref{sec:topoini}.
We show that the BEC order parameter decays exponentially in time.
On the other hand, observables that change the particle number by two have power-law decay.
The drastic difference arises because the latter observables have a local representation in terms of the Jordan-Wigner fermions while the former have a nonlocal string piece.
Section~\ref{sec:cfspin} uses the BEC quench results to make more quantitative estimates for the original spin problem.
We indeed find that the apparent persistent oscillation is due to the long lifetime as a result of low quasiparticle density, and we propose that one needs to simulate to longer time to see the decay.
In Sec.~\ref{sec:nonthermalizing}, we briefly comment on our study of the ``nonthermalizing" regime of Ref.~\cite{Banuls2011} using similar perturbative analysis and on its limitations.
We conclude in Sec.~\ref{sec:concl} with interesting outstanding questions raised by our work.

\section{Finite-size exact diagonalization comparison and identification of the oscillation frequency}
\label{finitesize}
To get some understanding of the observed weak thermalization behavior, we first study the same quench protocol as in Ref.~\cite{Banuls2011} using exact diagonalization (ED).
More specifically, we prepare the initial state as a product state where all spins are pointing in the $\hat{z}$ direction, $|Z\!+ \rangle$, and study its evolution under the Hamiltonian Eq.~(\ref{H}) with parameters $J = 1$ (taken as the energy unit), $h = 0.5$, and $g = -1.05$.
We consider a chain of length $L$ with periodic boundary conditions, $j + L \equiv j$.
Throughout, we set $\hbar = 1$.

\begin{figure}
    \includegraphics[width=0.95\columnwidth]{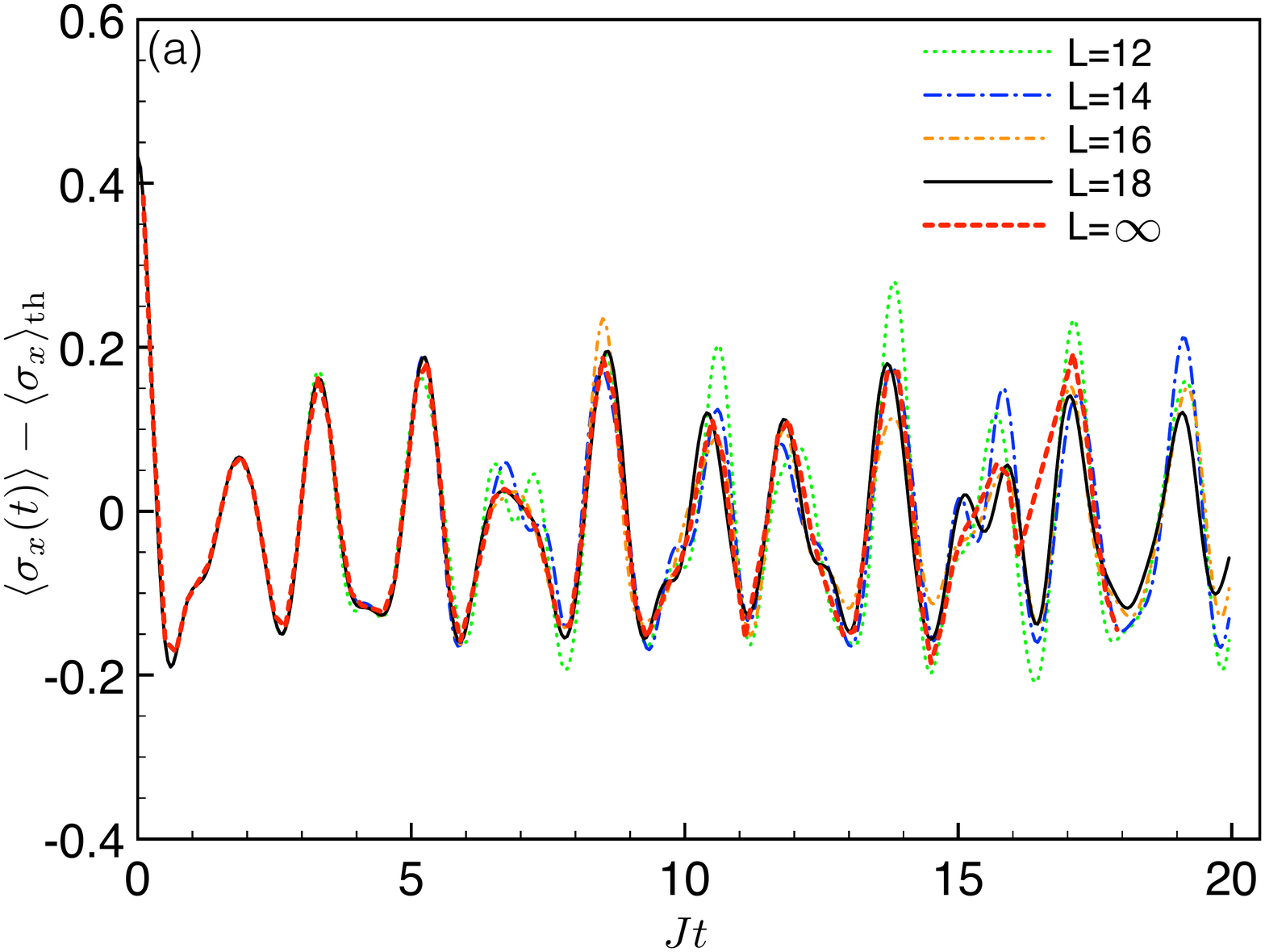}
    \includegraphics[width=0.95\columnwidth]{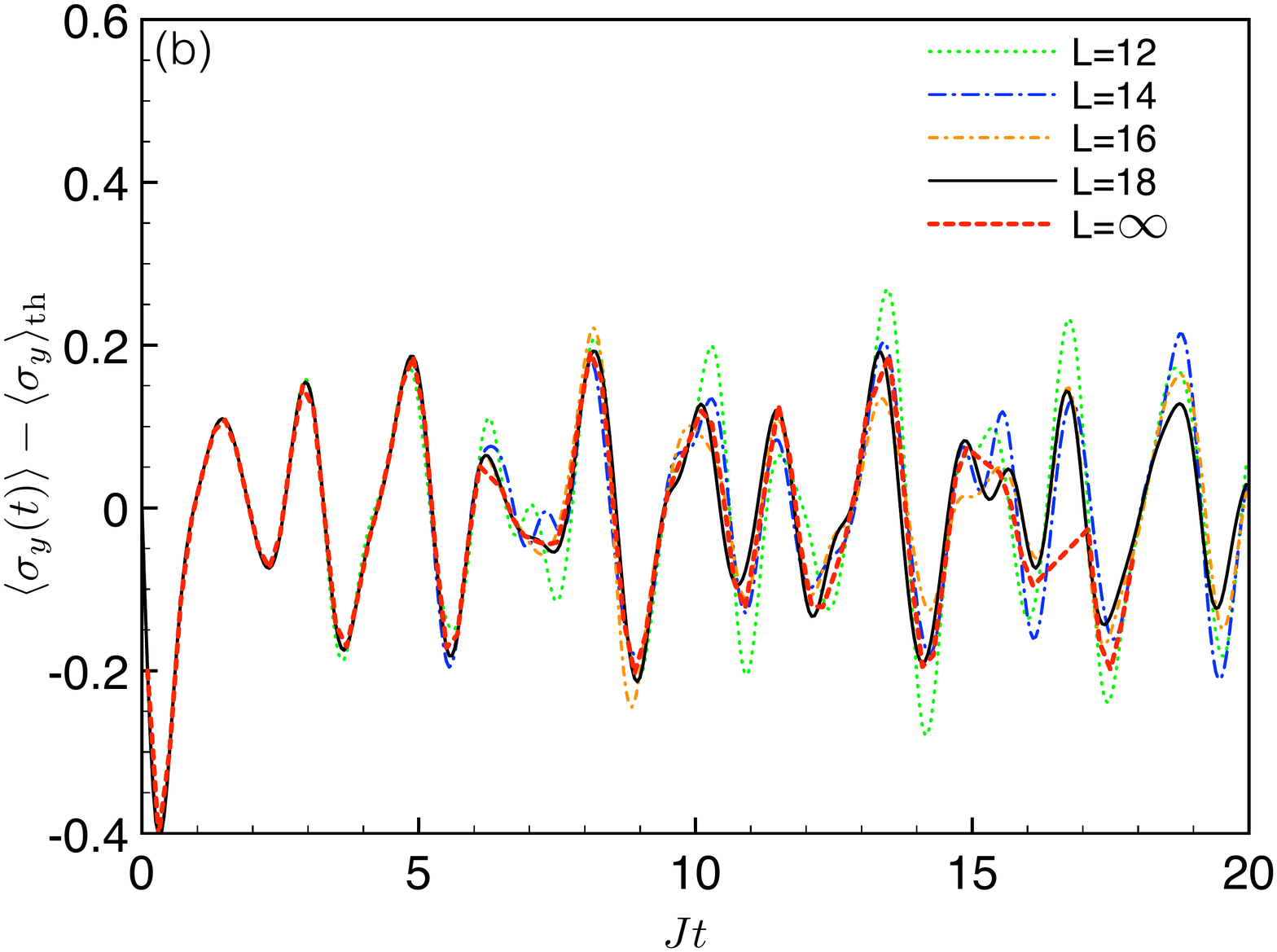}
    \includegraphics[width=0.95\columnwidth]{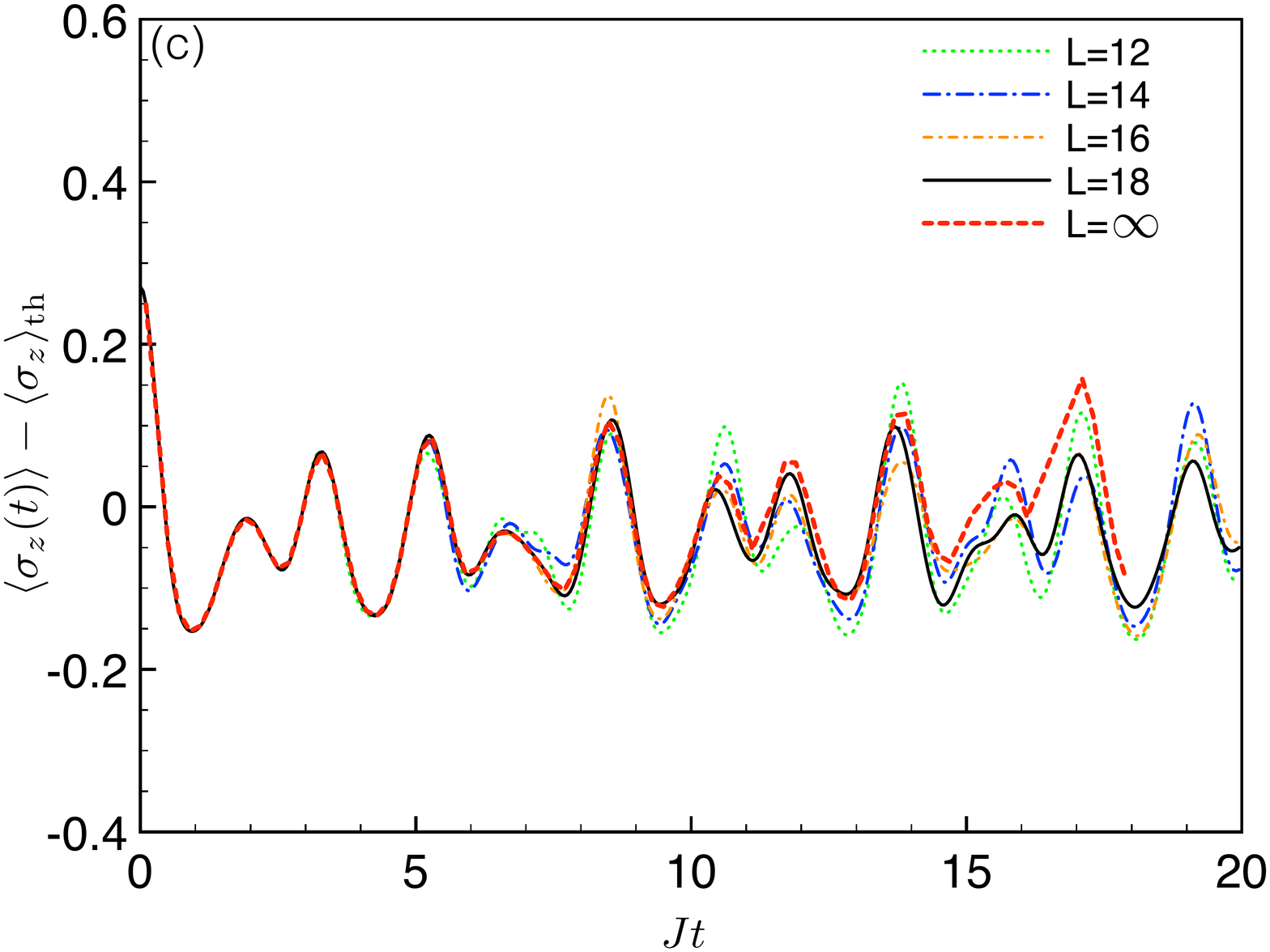}
    \caption{\label{fig:EDevolution} (color online)
(a)-(c) ED calculations of the evolution of some observables for several system sizes compared to infinite-MPS results from Ref.~\cite{Banuls2011} marked as $L = \infty$
(we are grateful to the authors of Ref.~\cite{Banuls2011} for sharing their data with us).
For smaller $L$, visible deviations from the $L = \infty$ results appear at smaller $t$, which we associate with recurrence phenomenon in finite systems.
Close agreement of our ED results with the infinite-MPS results over a large time window allows us to identify the frequency of the oscillations from finite-size spectra, which we argue is essentially the quasiparticle energy at zero momentum.}
\end{figure}

Figure~\ref{fig:EDevolution} shows comparisons of some local observables with the infinite-system results from Ref.~\cite{Banuls2011}.
Somewhat unexpectedly, our finite-size results for system size $L = 18$ capture the infinite-system results very closely up to time $t \approx 14$, which almost covers the full time window $t \leq 18$ displayed in Ref.~\cite{Banuls2011}.
By comparing ED results for a range of sizes between $L = 12$ and $L = 18$, we observe that the time $t_{\rm recurr}(L)$ beyond which the measurements deviate from the infinite-system results increases with the system size.
We expect that this time is roughly the time for the information to spread to the whole system, and beyond this time the ``recurrence'' phenomenon occurs.
For our largest size $L = 18$, the recurrence does not happen until $t_{\rm recurr} \approx 14$.

\begin{figure}
    \includegraphics[width=\columnwidth]{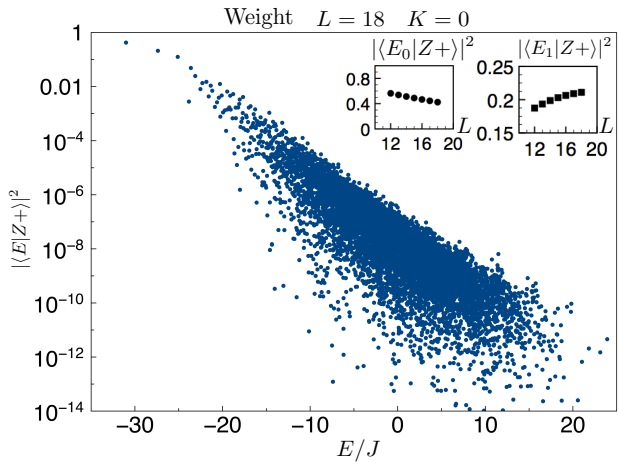}
    \caption{\label{fig:EDweights}
Weights of the initial state $|Z\!+ \rangle$ on the eigenstates $|E \rangle$ for our largest ED study with $L = 18$.
The figure only shows the weights on states with momentum quantum number $K = 0$, since the Hamiltonian is translationally invariant and the initial state has this quantum number (also, only states that are invariant under inversion have non-zero weights).
We see that even for our largest size, the majority of the weight is still on the ground state and the first excited state, which is expected since $\langle Z\!+ |H| Z\!+ \rangle = -1.5L = -27$ is close to $E_1$ for this size.
Insets: The system size dependence of the weights on the ground state $|E_0 \rangle$ and the first excited state $|E_1 \rangle$ from $L = 12$ to $L = 18$.
In the thermodynamic limit, the weights on these two states should go to zero, since $|Z\!+ \rangle$ has finite energy density above the ground state.
Nevertheless, $E_1 - E_0$ is defined in the thermodynamic limit and has a meaning of the quasiparticle gap, controlling oscillations of the observables as in Fig.~\ref{fig:EDevolution} over extended time interval.
}
\end{figure}

As a result of the close similarity between the ED and infinite-system results, we can understand the oscillation behavior from our ED spectra.
First of all, we observe that the oscillation frequency is essentially equal to the energy difference between the ground state and the first excited state.
In our calculation with $L = 18$, this energy difference is $\Delta E = E_1 - E_0 \approx 3.6401$.
We also note that even in our largest $L = 18$ system, the initial state actually has $|\langle \psi_{\rm ini} | \psi_0 \rangle|^2 \approx 42\%$ weight on the ground state and $|\langle \psi_{\rm ini} | \psi_1 \rangle|^2 \approx 21\%$ weight on the first excited state, as shown in Figure~\ref{fig:EDweights}.
Thus, one may think that the finite-size results are mainly determined by these large weights.

On the other hand, in an infinite system, the initial state has a finite energy density above the ground state.
Hence, the expansion of the initial state in terms of the eigenstates of the infinite system will also be dominated by eigenstates with finite energy density above the ground state.
In particular, the weights $|\langle \psi_{\rm ini} | \psi_0 \rangle|^2$ and $|\langle \psi_{\rm ini} | \psi_1 \rangle|^2$ will decay to zero exponentially in system size.
However, Ref.~\cite{Banuls2011} still found oscillations with apparently the same frequency in the infinite system.
A better picture is that the energy difference $\Delta E = E_1 - E_0$ in the finite-size system can be understood as a quasiparticle energy, which is defined also in the thermodynamic limit (in fact, $\Delta E$ is essentially converged to the quoted digits starting from $L = 10$).
The ground state is the vacuum of the quasiparticles, while the first excited state has one quasiparticle at momentum $k = 0$.
Therefore, the oscillation frequency in the finite-size system can be understood as the creation energy of the $k = 0$ quasiparticle.
Note that close to the ground state, any two states that differ by adding one such quasiparticle will have energy difference set by this quasiparticle energy.
If the corresponding matrix element for an observable is large and if the amplitudes of these states in $|\psi_{\rm ini} \rangle$ are significant, they will contribute to the observable with the same oscillation frequency.
Thus, this oscillation frequency is more robust than just the energy difference between the ground and first excited states in the finite system.
We can therefore infer that the quasiparticle excitation energy is the apparent oscillation frequency of the infinite-system calculation.
(As we will see in Sec.~\ref{sec:quenchMatter}, this is strictly true only in the low quasiparticle density limit, while in general the frequency will obtain density-dependent corrections.)

We can also make a rough estimate of the quasiparticle density in the initial state.
The average energy density is $\langle Z\!+| H |Z\!+ \rangle/L = -J - h = -1.5$.
With the quasiparticle energy $\Delta E = 3.6401$ and ground state energy density $E_0/L \approx -1.722$ (estimated from the $L = 18$ ED calculations and essentially converged in $L$ to the quoted digits), we can bound the quasiparticle density as $\rho \lesssim 0.061$.
This clearly demonstrates that the initial states in the weak thermalization regime in Ref.~\cite{Banuls2011} are states close to the low-energy part of the spectrum.
We can say that in this quasiparticle description, such initial states have low density of quasiparticles.
In this case, even though the spin model is a generic non-integrable model, the specific quench puts the system into a regime close to integrability in terms of the low-energy quasiparticles, which we believe is responsible for the observed weak thermalization behavior.

In statistical physics, we routinely calculate properties of many-body systems at low (but finite) temperatures by approximating the low-energy spectrum as a gas of non-interacting quasiparticles.
It is natural to ask if this picture can be used for studying quantum dynamics of states at low (but finite) energy density.
There is clearly some time scale over which the simple non-interacting picture gives sensible results, while here we want to focus on the asymptotic long-time behavior.
A common intuition is that in a generic non-integrable case, the residual interactions of the quasiparticles lead to eventual thermalization in the system, and this approach to thermalization can be studied by some semi-classical kinetic theory for weakly-interacting quasiparticles.
Our goal in the remainder of the paper is more modest.
We want to show that the oscillations in the above weak thermalization example eventually decay, using as much as possible only precise quantum mechanical arguments.
We will still be making some approximations but intuitively only in directions that make thermalization weaker, so our findings of the eventual decay under these approximations should translate to only stronger thermalization without the approximations.

\section{Perturbative picture of the quasiparticles and truncated Schrieffer-Wolff setup for the entire spectrum and quantum dynamics}
\label{perturbative}

\subsection{Low-energy quasiparticles}
\label{subsec:qppic}
To have a more precise formulation of the quasiparticle picture, we use a perturbative local Schrieffer-Wolff transformation \cite{MacDonald1988, Bravyi2011} near an exactly solvable limit where these quasiparticles are readily identified.
The corresponding SW-rotated picture can be viewed as an effective Hamiltonian for the quasiparticles, and we can also study the initial state and its evolution.  
We take
\begin{equation}
H_0 = -J \sum_{j=1}^L \sigma_j^z \sigma_{j+1}^z - h \sum_{j=1}^L \sigma_j^z
\label{H0}
\end{equation}
as our exactly solvable limit and treat
\begin{equation}
T = -g \sum_{j=1}^L \sigma_j^x
\label{T}
\end{equation}
as our perturbation.
This is not necessarily the best perturbative starting for our model parameters with sizable $g$ but will suffice for the qualitative picture.

$H_0$ is diagonal in the $\sigma^z$ basis.
Its energy levels are specified by the number $N_{\rm flip}$ of the spin-down sites, $\sigma_j^z = -1$, and the number $N_{\rm dw}$ of the domain walls, $\sigma_j^z \sigma_{j+1}^z = -1$.
The ground state has no spin-down sites and no domain walls, which is the $|Z\!+ \rangle$ state, while a state with $N_{\rm flip}, N_{\rm dw}$ has energy $2h N_{\rm flip} + 2J N_{\rm dw}$ above the ground state.
Note that $N_{\rm flip}$ and $N_{\rm dw}$ are not completely independent.
However, what is important later is that the number of different energy ``sectors" specified by $N_{\rm flip}, N_{\rm dw}$ is bounded by $L^2$ while the total number of states is growing as $2^L$.
Hence, many of the sectors are necessarily highly degenerate, particularly in the middle of the spectrum.
In cases where the density of spin-flips is small and they are well separated from each other, we can think of an isolated spin-flip as a quasiparticle with energy $2h + 4J$, but there are also quasiparticles with more structure.
Abusing the language somewhat, we will refer to the different $N_{\rm flip}, N_{\rm dw}$ sectors as having different quasiparticle numbers.

The action of the perturbation term $T$ changes the number of quasiparticle excitations and also introduces their dynamics.
The mixture of these effects is what makes the analysis very complicated.
To partially simplify the analysis, we find a unitary transformation $e^{iS}$ order by order to eliminate the effect of changing the excitation numbers, which gives us dynamical Hamiltonians that keep the number of quasiparticle excitations fixed (i.e., act within each sector $N_{\rm flip}, N_{\rm dw}$).
The detailed calculation is presented in App.~\ref{app:SW}.

To second order, we obtain the effective Hamiltonian as
\begin{equation}
H' = e^{iS} H e^{-iS} = H_0 + H_{\rm hop} + H_{\rm config} + H_{\rm other} ~,
\label{Hprime}
\end{equation} 
with
\begin{eqnarray*}
&& H_{\rm hop} = \left( \frac{-g^2}{2h} + \frac{g^2}{2h + 4J} \right) \sum_j P_{j-1} ^\up (\sigma_j^+ \sigma_{j+1}^- + \Hc) P_{j+2}^\up \\
&&~~ + \left( \frac{g^2}{2h} - \frac{g^2}{2h - 4J} \right) \sum_j P_{j-1}^\dn (\sigma_j^+ \sigma_{j+1}^- + \Hc) P_{j+2}^\dn ~, \\
&& H_{\rm config} = -\frac{g^2}{2h} \sum_j \left( P_{j-1}^\up \sigma_j^z P_{j+1}^\dn + P_{j-1}^\dn \sigma_j^z P_{j+1}^\up \right) \\
&&~~ -\frac{g^2}{2h + 4J} \sum_j P_{j-1}^\up \sigma_j^z P_{j+1}^\up 
- \frac{g^2}{2h - 4J} \sum_j P_{j-1}^\dn \sigma_j^z P_{j+1}^\dn ~,
\end{eqnarray*}
where $P_j^{\up, \dn} \equiv (1 \pm \sigma_j^z)/2$ are projectors to spin-up and spin-down states respectively and $\sigma_j^\pm \equiv (\sigma_j^x \pm i\sigma_j^y)/2$ are raising and lowering operators respectively.
The $H_{\rm hop}$ terms can be viewed as correlated hopping for the excitations.
As discussed previously, our initial state $|Z\!+ \rangle$ is close to the ground state, i.e., vacuum of quasiparticles.
Therefore, we expect the quasiparticles are effectively the down-spins, and the $H_{\rm hop}$ terms move such flipped spins, with additional dependencies on the neighboring spins.
The $H_{\rm config}$ terms describe additional contributions to the ``classical'' energy of the spin configuration, which can be viewed as some density-density-type interactions of the quasiparticles.
Note that the quasiparticles also have effective hard-core exclusion interaction.
As detailed in App.~\ref{app:SW}, $H_{\rm other}$ contains only contributions of order $O(g^3)$, including terms that preserve the excitation numbers and also terms that change the excitation numbers.
We will make an approximation where we drop the $H_{\rm other}$ terms (more discussion below) and call the resulting $H_{\rm eff}$ as ``effective Hamiltonian," which acts separately in each sector.
Thus, by eliminating the leading excitation-number-changing effect in our original Hamiltonian and dropping the $H_{\rm other}$ terms, the dynamics now can be roughly viewed as hard-core bosons with hopping and interaction in the dilute limit, where $\sigma_j^z = -1 (+1)$ corresponds to the presence (absence) of bosons.

\begin{figure}
    \includegraphics[width=\columnwidth]{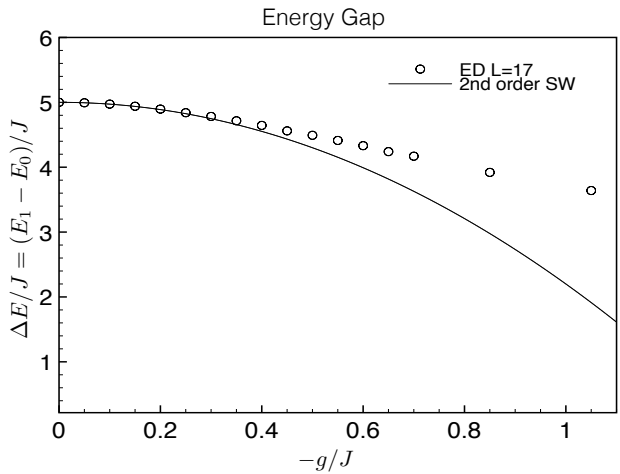}
    \caption{\label{fig:massgap}
The energy difference between the first excited state and the ground state as a function of $g$.  
This gap can also be understood as quasiparticle excitation energy.
The ED result is from system size $L = 17$ (the ED gap estimates are essentially already converged for $L \gtrsim 10$).
We also show perturbative SW result at order $O(g^2)$, Eq.~(\ref{particleMass}).
The exact and perturbative calculations agree well at small $g$ and deviate more at larger $g$, but the qualitative picture of the quasiparticle is robust since the gap does not close over the range of $g$ shown.}
\end{figure}

For excitations that are widely separated spin-flips, we need to consider only the first term in $H_{\rm hop}$ and the first and second terms in $H_{\rm config}$.
The single quasiparticle energy at momentum $k$ is readily evaluated as
\begin{equation}
\epsilon_k = 2h + 4J - \frac{g^2}{h} + \frac{2g^2}{h + 2J}
-\left( \frac{g^2}{2h} - \frac{g^2}{2h + 4J} \right) 2 \cos(k) ~.
\label{qpdispersion}
\end{equation}
The quasiparticle energy at zero momentum, which is relevant for the oscillations in the quench problem of interest, is
\begin{equation}
\label{particleMass}
\epsilon_{k=0} = 2h + 4J - \frac{2g^2}{h} + \frac{3g^2}{h + 2J} ~.
\end{equation}
Figure~\ref{fig:massgap} shows the excitation energy as a function of $g$ in the range between $0$ and $-1.05$.
The perturbative calculation is accurate for small $|g| \lesssim 0.25$ but becomes less accurate at larger $|g|$.
In particular, this second-order calculation would give $\epsilon_{k=0} \approx 1.913$ at $g = -1.05$, while the true gap $\Delta E = 3.6401$ is almost two times larger.
This quantitative difference is not surprising given that the assumption $g \ll h, J$ is clearly not satisfied in this case (particularly since energy denominators $2h$ appear when treating the sector with one quasiparticle perturbatively in $g$).
However, the true gap remains large in this range of $g$, and the quasiparticle picture developed perturbatively in $g$ remains qualitatively correct (it can be further improved if needed, but the presented picture is already sufficient for our discussion).

By examining the low-energy ED spectra for $g = -1.05$, we find that the band of states closest to the ground state is well-described by $\epsilon_k = \epsilon_{k=0} + 2 J_{\rm eff} [1 - \cos(k)]$ with $J_{\rm eff} \approx 0.44$.
The effective hopping amplitude again differs quantitatively from the perturbative estimate in Eq.~(\ref{qpdispersion}), but our overall picture of the quasiparticles at low energy is robust.

We finally note that the picture of weakly-interacting spin-flips is not accurate here.
Two spin flips can lower their energy by roughly $4J$ if they are next to each other, i.e., there is a significant attractive interaction between them.
In the SW perturbative treatment, the next sector in energy after the single-spin-flip sector ($N_{\rm flip} = 1, N_{\rm dw} = 2$) has two flipped spins that are next to each other ($N_{\rm flip} = 2, N_{\rm dw} = 2$) but otherwise can be anywhere on the chain.
The effective Hamiltonian to second order in $g$ gives energy $4h + 4J + 2g^2/(h+2J)$ and no dispersion for these states, while of course some dispersion will develop at higher order.
In fact, by examining the low-energy ED spectra for $g = -1.05$, we find the single-spin-flip band of $L$ states covering energy window $[\Delta E, 5.31]$ above the ground state ($\Delta E = 3.6401$ is the gap), and then another band of $L$ states covering energy $[5.90, 6.84]$ and separated from the next set of states starting at $\approx 2\Delta E = 7.28$.
The second band can be viewed as corresponding to a stable bound state of two spin-flips, which is hopping around with an amplitude about two times smaller than the single spin-flip.
On the other hand, the states above $2\Delta E$ can be viewed as corresponding to the two-spin-flip continuum with well-separated spin-flips.
We can in principle view the bound state as another quasiparticle in the system at low energy and now think about dilute gas of these as well as single-spin-flip quasiparticles, adding more accuracy to the description but also much more complexity.
However, we will not use such details below and will proceed with a more crude picture and language of quasiparticles as if they were only single-spin-flips.
This is an OK approximation at low energy density but can become quantitatively inaccurate at somewhat higher density.

\subsection{Truncated Schrieffer-Wolff transformation for dynamics} 
\label{truncatedSW}
While our original motivation for using the Schrieffer-Wolff transformations was to understand the low-energy quasiparticles, the transformation as defined acts on the entire Hamiltonian and the entire spectrum.
We can boldly try to use the rotated $H'$ with the $H_{\rm other}$ terms omitted and study the quantum dynamics under this ``effective Hamiltonian.''
By doing so, we are essentially postulating an emergent integral of motion, namely the quasiparticle number conservation, or more precisely, preservation of the sector identities.
Recent works Refs.~\cite{Grover2014, Garrison2016} conjectured possible emergence of such integrals of motion in translationally invariant systems as a (much weaker) analog of many-body localization physics without underlying disorder.
However, this conjecture is far from being established, and we will not try to prove or disprove it here.
If such an emergence of the new integral of motion were true, this would likely mean absence of full thermalization in the present context, as one would then expect ``equilibration" to an appropriate generalized Gibbs ensemble treating the new integral of motion.
Nevertheless, we will see that even in this case the oscillations in the observables still decay, i.e., the weak thermalization turns to a more conventional thermalization at long times.
If the conjecture is not true, then our calculations in the truncated SW scheme can be viewed as providing sufficient mechanisms for thermalization, while in the full picture without the new integral of motion the thermalization is likely to proceed only faster.

Keeping the above remarks in mind, we now describe calculations in the truncated SW-rotated picture.
The time evolution of an observable $\hat{O}$ becomes 
$\langle \psi_{\rm ini} | e^{iHt} \hat{O} e^{-iHt} |\psi_{\rm ini} \rangle = \langle \psi_{\rm ini}' | e^{iH't} \hat{O'} e^{-iH't} |\psi_{\rm ini}' \rangle$,
where $\hat{O'} = e^{iS} \hat{O} e^{-iS}$ and $|\psi_{\rm ini}' \rangle = e^{iS} |\psi_{\rm ini} \rangle$ are the appropriately rotated operator and initial state.

Consider first observables in the rotated picture.
For the observables that we study,
\begin{eqnarray}
&& \left( \sigma_j^x \right)' \approx \sigma_j^x + O(g) ~, \\
&& \left( \sigma_j^y \right)' \approx \sigma_j^y + O(g) ~, \\
&& \left( \sigma_j^z \right)' \approx \sigma_j^z - \frac{g}{h} \left( P_{j-1}^\up \sigma_j^x P_{j+1}^\dn + P_{j-1}^\dn \sigma_j^x P_{j+1}^\up \right) \nonumber \\
&& - \frac{g}{h + 2J} P_{j-1}^\up \sigma_j^x P_{j+1}^\up - \frac{g}{h - 2J} P_{j-1}^\dn \sigma_j^x P_{j+1}^\dn + O(g^2).~~~
\label{sigmazprime}
\end{eqnarray}
Of course, $\left( \sigma_j^x \right)'$ has an $O(1)$ component onto the operator $\sigma_j^x$ that changes the quasiparticle number by one and hence ``detects'' the quasiparticle energy, and similarly for $\left( \sigma_j^y \right)'$.
On the other hand, the leading contribution to $\left( \sigma_j^z \right)'$ does not change the quasiparticle number.
However, we can see that in the rotated picture at order $O(g)$, this observable also contains $\sigma_j^x$, which detects the quasiparticle energy.
The above expressions explain why the oscillations in $\langle \sigma_j^x(t) \rangle$ and $\langle \sigma_j^y(t) \rangle$ in Fig.~\ref{fig:EDevolution} have roughly similar amplitudes but are shifted in phase by $\pi/2$, while the oscillation in $\langle \sigma_j^z(t) \rangle$ has a smaller amplitude and is in phase with $\langle \sigma_j^x(t) \rangle$ [indeed, the dominant term in $\left( \sigma_j^z \right)'$ in the regime of low quasiparticle density is $-\frac{g}{h + 2J} P_{j-1}^\up \sigma_j^x P_{j+1}^\up$ and $g < 0$.]
Thus, our quasiparticle picture of the origin of oscillations can explain even finer details in the numerical results.
Finally, we note that operators $\left( \sigma_j^x \right)'$ and $\left( \sigma_j^y \right)'$ at next order contain contributions that create two spin-flips [see Eq.~(\ref{sigmaxprime}) and Eq.~(\ref{sigmayprime}) for explicit formulas].
Hence, when discussing observables in the rotated SW picture, we should also consider operators that change the excitation number by two.

Consider now the initial state in the rotated picture.
Since $iS$ is a local operator containing spin-flip terms (see App.~\ref{app:SW} for explicit formulas), we can think of $|\psi_{\rm ini}' \rangle = e^{iS} |\psi_{\rm ini} \rangle$ roughly as a product state where the spin on each site is rotated a little away from the $\hat{z}$-direction.
In terms of hard-core bosons representing the spin-flips (quasiparticles), this state of course has some small density of bosons, since $n_j = (1 - \sigma_j^z)/2$.
More crucially, it is actually a Bose-Einstein condensate (BEC) state, since the rotated spin state can be written in the boson language as, approximately, $\prod_j (\alpha + \beta b_j^\dagger) |0 \rangle$, where $b_j^\dagger \equiv (\sigma_j^x - i \sigma_j^y)/2$ (denoted $\sigma_j^-$ earlier and in App.~\ref{app:SW}). 

\begin{figure}
   \includegraphics[width=\columnwidth]{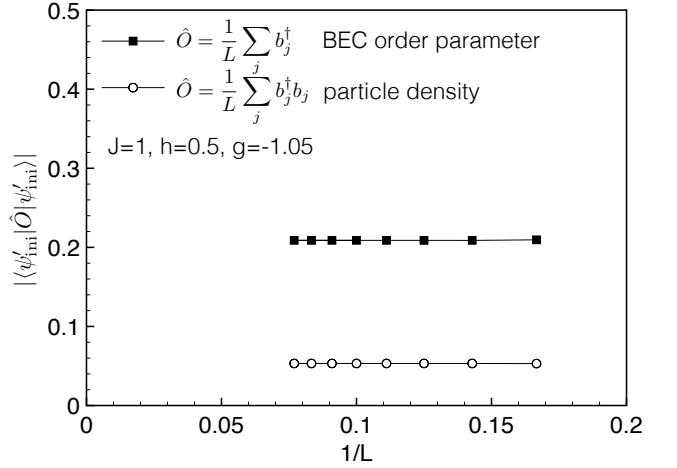}
   \caption{\label{fig:InitialState}
Properties of the SW-rotated initial state $|\psi_{\rm ini}' \rangle = e^{iS} |\psi_{\rm ini} \rangle$, with $iS = iS^{[1]} + iS^{[2]}$ calculated to second order in $g$ (see App.~\ref{app:SW} for details).
The figure shows expectation values of the particle density $b_j^\dagger b_j \equiv (1 - \sigma_j^z)/2$ and BEC order parameter $b_j^\dagger \equiv (\sigma_j^x - i \sigma_j^y)/2$.
The particle density $\rho \approx 0.05$ is very low and close to the estimate based on the average energy density in the initial state and the quasiparticle gap.}
\end{figure}

To get a more quantitative characterization of the initial state in the rotated picture, we calculated $|\psi_{\rm ini}' \rangle = e^{iS} |\psi_{\rm ini} \rangle$ for system sizes $L = 6$ to $L = 13$, using $iS$ calculated to second order in $g$ from App.~\ref{app:SW} and applying true unitary $e^{iS}$.
Figure~\ref{fig:InitialState} shows measurements of the boson density and also of the BEC order parameter as a function of inverse system size $1/L$.
The values are essentially converged in the first four non-zero digits.
We can see that the density is roughly $\rho \approx 0.05$, consistent with our previous estimate and our picture of diluteness of quasiparticles.
Furthermore, the initial state indeed has nonzero BEC order parameter.

\begin{figure}
\includegraphics[width=\columnwidth]{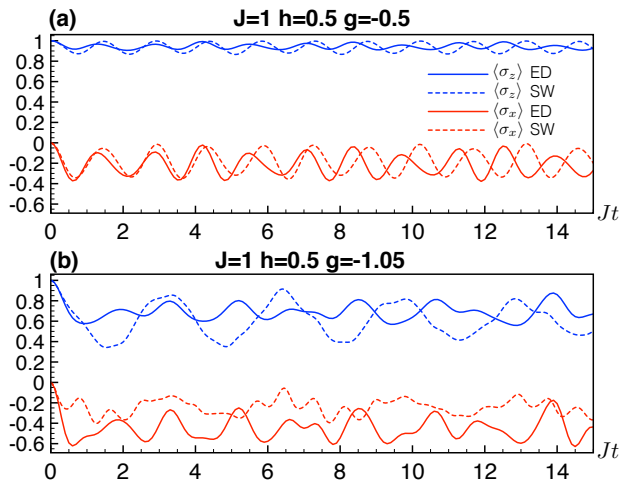}
   \caption{\label{fig:SWvsED}(color online) Comparisons of the dynamics from ED and 2\textsuperscript{nd}-order SW for two different parameters for system size $L=13$. (a)~$g=-0.5$. The perturbative calculation is still quantitatively accurate at short time. The difference between ED and SW mostly comes from the difference between the exact and perturbative quasiparticle energy gap, resulting in the different oscillation frequencies. (b)~$g=-1.05$. The quantitative comparison becomes inaccurate at this parameter. For example, the oscillation frequency in the SW calculation is almost half of the ED calculation. Nevertheless, the SW calculation still captures the qualitative behavior observed in the ED calculation.     
}
\end{figure}

To gain some intuition about the truncated SW picture, we performed numerical calculations in the truncated SW picture as follows.
For more accuracy, we start with the rotated state $|\psi_{\rm ini}' \rangle = e^{iS} |\psi_{\rm ini} \rangle$ and observables $\hat{O'} = e^{iS} \hat{O} e^{-iS}$ using true unitary $e^{iS}$ with $iS = iS^{[1]} + iS^{[2]}$ calculated to second order in $g$.
Note that an exact calculation of $e^{iS}$ is possible on small sizes.
However, for the dynamical Hamiltonian, we use the perturbatively developed $H'$ omitting $H_{\rm other}$ terms.
Of course, if we used exactly-rotated $H' = e^{iS} H e^{-iS}$, everything would be identical to the original calculation with the un-rotated initial state, observables, and Hamiltonian, while the setup where we use the truncated $H'$ allows us to gauge the effect of the truncation.
In principle, the effective Hamiltonian completely separates the different energy scales of the original problem.
Thus, the ``large" energy scales $h$ and $J$ determine only the spacing between the sectors and would enter only the frequency of the oscillations, while all the processes inside each sector---kinetic energy, effect of hard-core exclusion, and explicit interactions---are now controlled by one energy scale $O(g^2)$ (here for the sake of simplicity we ignore the difference between effects of $h$ and $J$ and imagine them as giving one energy scale).

Figure~\ref{fig:SWvsED} compares the 2\textsuperscript{nd}-order SW calculation described in the previous paragraph with the ED result, for system size $L=13$ and different parameters. We find persistent oscillations on the time scales similar to those in Fig.~\ref{fig:EDevolution}. As shown in the figure, for $g=-0.5$, the perturbative description is still roughly quantitatively accurate. The oscillations are somewhat more regular, consistent with the expectation that the truncation reduces decoherence effects. 
The main difference is due to the frequency difference, as the perturbative calculation of the quasiparticle gap has a small error compare to the ED gap. On the other hand, for $g=-1.05$, the SW calculation deviates significantly from the ED calculation, which is not surprising, since the parameter $g$ is not small anymore. 
Nevertheless, the ED result of $g=-1.05$ is still qualitatively similar to the ED result of $g=-0.5$, i.e., both have persistent oscillation with frequency given by the quasiparticle energy gap.
Moreover, the quasiparticle gap has not closed at $g=-1.05$ as shown in Fig.~\ref{fig:massgap}. 
Therefore, we conjecture that the truncated Hamiltonian obtained from SW transformation is still suitable to describe the dynamics, however, with the parameters understood as the renormalized values instead of values calculated from the perturbative formula directly. 

To conclude the above discussion, the strong oscillation behavior is mainly coming from the measurement of $\langle \psi_{\rm ini}' | b_j^\dagger(t) | \psi_{\rm ini}' \rangle$ in the boson language.
Effectively, this is the evolution of the BEC order parameter with the dynamical Hamiltonian of interacting bosons and the initial BEC state as we discussed.
The fate of the oscillations is not an entirely trivial problem, as can be seen from the following considerations.
In fact, in the extremely simplified case where the initial state has BEC and the dynamical Hamiltonian is purely boson hopping $H = -J \sum_j (B_j^\dagger B_{j+1} + \Hc) + W \sum_j N_j = \sum_k [W - 2J \cos(k)] B_k^\dagger B_k$ without interaction and without hard-core constraint, the evolution indeed exhibits undamped oscillation with frequency $\omega = W - 2 J$.
Note that here $B_j$ are canonical (not hard-core) bosons and $N_j = B_j^\dagger B_j$ (we used capital letters to distinguish from hard-core bosons used in the next section).
Furthermore, allowing interactions among quasiparticles of the type typically done in the Landau's Fermi liquid theory, $H_{\rm int} = 1/(2L) \sum_{k,p} V_{k,p} N_k N_p$, leads only to shifting the oscillation frequency by an amount $(1/L) \sum_p V_{0,p} \langle N_p \rangle$, 
where $\langle N_p \rangle$ is the expectation value in the initial state, but not to decay of the oscillations at long times.
Only when we allow more general interactions, we expect that the BEC order parameter will start to damp.
In the next section, we will show that already the hard-core exclusion will lead to decrease of the oscillations at long times.

\section{Quench of BEC state to solvable hard-core boson Hamiltonian}
\label{sec:quenchMatter}
As discussed in the previous section, the dynamics of the quantum spin chain after removing the excitation-changing part can be viewed as an interacting hard-core boson problem.
Even though we can obtain the quasiparticle description, the effective problem is still very difficult to analyze due to its non-trivial interactions.
Therefore, we further simplify the problem by considering more simple initial states and a solvable effective Hamiltonian.

Specifically, we consider the dynamical Hamiltonian
\begin{equation}
\label{Hhcb}
H = W_{\rm eff} \sum_{j=1}^L b_j^\dagger b_j - J_{\rm eff} \sum_{j=1}^L \left(b_j^\dagger b_{j+1} + \Hc \right) ~,
\end{equation}
with the hard-core constraint $(b_j^\dagger)^2 = 0$.
We also consider periodic boundary conditions $b_{j+L} \equiv b_j$ to be closer to the thermodynamic limit.
This Hamiltonian can be viewed as an approximation to the effective Hamiltonian in Eq.~(\ref{Hprime}) where we drop $H_{\rm other}$ and two-site and three-site interaction terms.
Note that the parameter $J_{\rm eff}$ here is not related to the spin interactions in the original spin chain but should be viewed instead as the effective hopping amplitude of the spin-flips in $H_{\rm hop}$; this should not cause any confusion since in this section we will focus on the above hard-core boson model.
We will show that even dropping these interaction terms, the BEC order parameter will still decay.
We would expect that including the dropped terms would allow more channels for thermalization, although details of the interactions can certainly have quantitative effects. 
We will discuss this approximation and  the effects of additional interactions in Sec.~\ref{sec:cfspin}.

The advantage of the above simplified Hamiltonian is that it is exactly solvable.
Using Jordan-Wigner (JW) transformation, which transforms the hard-core bosons to fermions,
\begin{equation}
\label{JWtransform}
b_j = \left( \prod_{j'=1}^{j-1} e^{i \pi n_{j'}} \right) c_j ~,
\end{equation} 
the Hamiltonian can be rewritten in the fermionic representation as 
\begin{eqnarray}
H &=& W_{\rm eff} \sum_{j=1}^L c_j^{\dagger} c_j - J_{\rm eff} \sum_{j=1}^{L-1} \left(c_j^\dagger c_{j+1} + \Hc \right) \\
&-& J_{\rm eff} (-1)^{N_{\rm tot} + 1} \left(c_L^\dagger c_1 + \Hc \right) ~,
\end{eqnarray}
where $N_{\rm tot} \equiv \sum_{j=1}^L n_j$ is the total particle number.
As is well-known, in the fermionic representation, for sectors with even $N_{\rm tot}$ we effectively have anti-periodic boundary conditions, while for sectors with odd $N_{\rm tot}$ we have periodic boundary conditions.
We can then use Fourier transformation $c_k = (1/\sqrt{L}) \sum_j c_j e^{-i k j}$ to diagonalize the Hamiltonian $H = \sum_k \epsilon_k c_k^\dagger c_k$, where $k = 2\pi (m + 1/2)/L$ for even-particle-number sectors and $k = 2\pi m/L$ for odd-particle-number sectors, with $m=0, 1, \dots, L-1$.
The single-particle dispersion is $\epsilon_k = W_{\rm eff} - 2 J_{\rm eff} \cos(k)$.

The difference in the boundary conditions for the even and odd sectors is not important when considering the spectrum and static properties in the thermodynamic limit.
However, when considering the dynamics of observables connecting different number-parity sectors, ignoring the boundary conditions and the resulting differences in $c_k$ used to diagonalize the even and odd sectors (ultimately related to the string operator when connecting such sectors), results in an erroneous answer, as we will explicitly show below.
This is also the major obstacle to obtaining analytical results for $\langle \psi_{\rm ini}| b_j^\dagger(t) | \psi_{\rm ini} \rangle$, with any reasonable initial state $|\psi_{\rm ini} \rangle$.
On the other hand, we can obtain a compact analytical expression for $\langle \psi_{\rm ini}| b_j^\dagger(t) b_{j+1}^\dagger(t) |\psi_{\rm ini} \rangle$, which connects sectors with the same number parity.
The reason is that the Heisenberg representation of $c_k(t) = c_k e^{-i \epsilon_k t}$ only makes sense when constructing operators that connect sectors with the same number parity.

In the next two subsections, we will consider two different initial states, both with nonzero boson condensation, evolving under Hamiltonian Eq.~(\ref{Hhcb}).
For $W_{\rm eff} > 2 J_{\rm eff}$, the ground state of this Hamiltonian in the full Fock space is a trivial Mott insulator.
However, this and the value of $W_{\rm eff}$ are actually not important for the relaxation dynamics:
One can readily see that $W_{\rm eff}$ only adds to the oscillation frequency for processes connecting sectors with different $N_{\rm tot}$ and not to any relaxation dynamics.
The latter is determined by the spectra inside each sector, where $J_{\rm eff}$ is the only energy scale.
Of course, the properties of the initial state are also important (e.g., which $N_{\rm tot}$ are present, the energy distribution, etc.).
For illustration, we will take $W_{\rm eff} = 5 J_{\rm eff}$ to see multiple oscillations before recurrence time.
Given our motivation for this model as a simplified effective model for quasiparticles in the original spin problem in the regime of low quasiparticle density, we would like to consider initial states with low average particle density.
We will first consider higher densities to better see qualitative behaviors, and afterwards we will return to more appropriate parameters for the original motivation.
We will consider quantities $\langle \psi_{\rm ini}| b_j^\dagger(t) |\psi_{\rm ini} \rangle$ (the matter wave or the BEC order parameter) and $\langle \psi_{\rm ini}| b_j^\dagger(t) b_{j+1}^\dagger(t) |\psi_{\rm ini} \rangle$ (pair-boson condensation order parameter).
One motivation for considering both single- and pair-boson operators comes from the fact that both these generically contribute to the observables of interest in the rotated SW picture for the original spin problem, see discussion after Eq.~(\ref{sigmazprime}) and Eqs.~(\ref{sigmaxprime})-(\ref{sigmayprime}) in App.~\ref{app:SW}.
An even stronger motivation comes from intrinsic interest in the integrable hard-core boson model, as comparing these quantities will illustrate the importance of the boundary conditions on the Jordan-Wigner fermions and ultimately of the string operator when connecting sectors with different number parity.

\subsection{Initial hard-core boson BEC state as a product state}
\label{productini}
We first consider our initial state as a hard-core boson coherent state
\begin{equation}
\label{hcbBEC}
|\psi_{\rm ini,A} \rangle = \prod_{j=1}^L (\alpha + \beta b_j^\dagger) |0 \rangle ~,
\end{equation}
where $|0 \rangle$ is the vacuum of the hard-core bosons.
The normalization requires $|\alpha|^2 + |\beta|^2 = 1$.
The boson density is $\rho = \langle \psi_{\rm ini,A}| b_j^\dagger b_j |\psi_{\rm ini,A} \rangle = |\beta|^2$, while the BEC order parameter is $\Phi = \langle \psi_{\rm ini,A}| b_j^\dagger |\psi_{\rm ini,A} \rangle = \beta^* \alpha$.

\begin{figure}
    \includegraphics[width=\columnwidth]{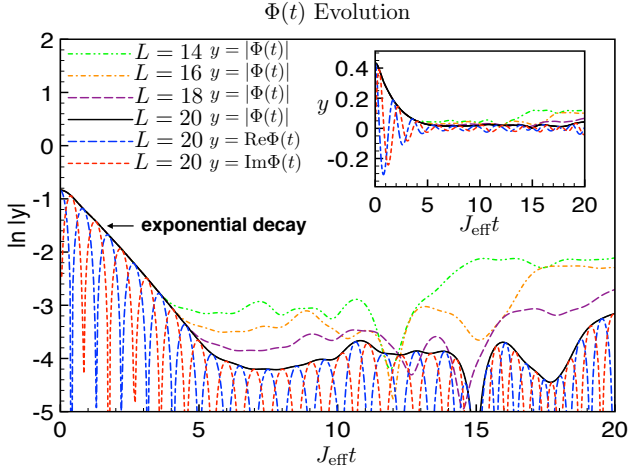}
    \caption{\label{bevol} (color online)
ED results for $\Phi(t) \equiv \langle b_j^\dagger(t) \rangle$ for the hard-core boson model, Eq.~(\ref{Hhcb}), and the product BEC state, Eq.~(\ref{hcbBEC}), with average boson density $\rho = |\beta|^2 = 0.25$, for system sizes $L = 14, 16, 18, 20$.
The parameters of the dynamical Hamiltonian are $J_{\rm eff} = 1$ and $W_{\rm eff} = 5 J_{\rm eff}$.
Note that the main panel shows $\ln |\Phi(t)|$ vs $J_{\text{eff}}t$, and the results strongly suggest exponential decay in $t$ until recurrence time: as the system size increases, the time interval over which the exponential decay is observed also increases.
Inset: linear scale for the observable.
The recurrence times where the smaller-size results peel off from the largest-size results are roughly in agreement with those in Fig.~\ref{bbevol}.}
\end{figure}

\begin{figure}
    \includegraphics[width=\columnwidth]{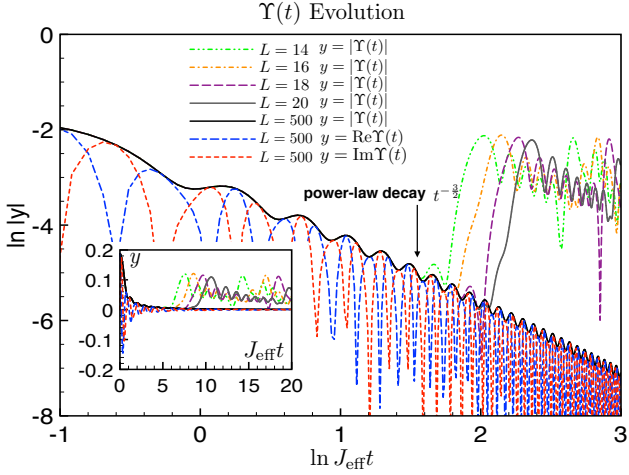}
    \caption{\label{bbevol} (color online)
ED results for $\Upsilon(t) \equiv \langle b_j^\dagger(t) b_{j+1}^\dagger(t) \rangle$ for the same systems as in Fig.~\ref{bevol} with average density $\rho = 0.25$.
This observable can be also calculated analytically using Jordan-Wigner transformation (checked against ED for small $L$), allowing us to study much larger sizes and times, as illustrated here for $L = 500$.
Note that the main panel shows log-log plot, and the long-time behavior in the largest system clearly has a power-law envelope with decay $t^{-3/2}$, in agreement with the analytical calculation in the text.
Inset: linear scale for the observable.
By comparing data for $L = 14, \dots, 20$ and much larger $L = 500$, we can also clearly see where the recurrences appear for the smaller sizes.}
\end{figure}

Figure~\ref{bevol} shows ED results for the evolution of the BEC order parameter
\begin{equation}
\Phi(t) \equiv \langle b_j^\dagger(t) \rangle \equiv \langle \psi_{\rm ini,A}| b_j^\dagger(t) |\psi_{\rm ini,A} \rangle
\end{equation}
for the initial state with average density $\rho = 0.25$ and system sizes up to $L = 20$.
From the semi-log plot, it is clear that the initial decay is exponential.
At later times, the ED results again suffer from the finite-size recurrence effect.
However, as the system size increases, we see the exponential decay over a longer time interval.
Therefore, we infer that in the thermodynamic limit the BEC order parameter decays exponentially in time even in this integrable system.

We remark that a naive calculation that would ignore the different boundary conditions for the Jordan-Wigner fermions in different number-parity sectors would suggest a different (wrong) result.
By translational invariance, the BEC order parameter is just $\langle \psi_{\rm ini,A}| b_{j=1}^\dagger(t) |\psi_{\rm ini,A} \rangle$, and one could naively proceed
\begin{eqnarray}
&& \langle \psi_{\rm ini,A}| b_{j=1}^\dagger(t) |\psi_{\rm ini,A} \rangle
= \langle \psi_{\rm ini,A}| c_{j=1}^\dagger(t) |\psi_{\rm ini,A} \rangle \nonumber \\
&& \mathrel{\stackrel{\makebox[0pt]{\mbox{\normalfont\tiny !!! wrong !!!}}}{=}} ~~~~
\frac{1}{\sqrt{L}} \sum_k \langle \psi_{\rm ini,A}| c_k^\dagger |\psi_{\rm ini,A} \rangle e^{i \epsilon_k t} e^{-i k}\\
&& = \frac{1}{L} \sum_k \sum_j \langle \psi_{\rm ini,A}| c_j^\dagger |\psi_{\rm ini,A} \rangle e^{i \epsilon_k t} e^{i k j} e^{-i k} ~.   
\end{eqnarray}
Expectation values $\langle \psi_{\rm ini,A}| c_j^\dagger |\psi_{\rm ini,A} \rangle$ can be easily evaluated in the product BEC state.
From here, the calculation is not sensitive to the details of the sum over $k$, which can be turned into an integral for large $L$, and a standard steepest descent analysis of the last equation would give $t^{-\frac{1}{2}}$ decay at large time $t$.
However, this calculation is wrong at the emphasized step, since there is no well-defined fermionic quasiparticle creation operator $c_k^\dagger$ acting between sectors with different parity.
Ultimately, this is related to the non-local character of the boson order parameter in terms of the JW fermions, which in the translationally invariant case with periodic boundary conditions for the bosons yields effectively different boundary conditions for the JW fermions in the even and odd sectors.
(In the case with open boundary conditions, the above calculation focusing on the site at the left boundary would not be representative of the infinite system, while a valid calculation with a site in the middle of the system would have to time-evolve with the string operator.)
In this case, we do not have a simple analytical calculation of the observable even though the model is solvable by JW fermions.
A more involved analytical calculation supporting exponential decay of $\Phi(t)$ will be presented in the next subsection.

On the other hand, Fig.~\ref{bbevol} shows the evolution of 
\begin{equation}
\Upsilon(t) \equiv \langle b_j^\dagger(t) b_{j+1}^\dagger(t) \rangle \equiv \langle \psi_{\rm ini,A}| b_j^\dagger(t) b_{j+1}^\dagger(t) |\psi_{\rm ini,A} \rangle ~.
\end{equation}
We actually have a full analytical calculation of this observable and a closed-form expression in the thermodynamic limit.
Nevertheless, we still show the finite-size ED results (which we also checked against the analytical calculations), as a reference to compare with the results for $\Phi(t)$.
We can see that this pair-boson observable decays with a power-law envelope $t^{-3/2}$ until the recurrence phenomenon sets in.
Again, as we increase the system size, the recurrence time also increases.

We present the analytical calculation in App.~\ref{app:BB}, while here we only show the final result in the thermodynamic limit,
\begin{equation}
\label{bbeq}
\Upsilon(t) = 2(\beta^* \alpha)^2 \int_{-\pi}^\pi \frac{dk}{2\pi}~ \frac{\sin^2(k)}{1 + \eta^2 - 2\eta\cos(k)} e^{2 i \epsilon_k t} ~,
\end{equation}
where $\eta \equiv |\alpha|^2 - |\beta|^2$.
The long-time behavior is controlled by extrema of $\epsilon_k$; these occur at $k = 0$ and $k = \pi$, so we expect oscillations at two frequencies, $2 \epsilon_{k = 0/\pi} = 2(W \mp 2J)$.
Since in the integrand the factor multiplying $e^{2 i \epsilon_k t}$ vanishes at both these points, a saddle-point analysis gives power-law envelope $t^{-3/2}$. Both these frequencies and the power-law envelope are indeed observed in the numerical calculations in Fig.~\ref{bbevol}.

Comparing our numerical results for $\Phi(t)$ and $\Upsilon(t)$, we can confidently say that the latter observable decays more slowly, despite being a composite operator in terms of microscopic bosons.
On the time scales where the behavior is representative of the thermodynamic limit, the former observable decays faster than power law and is consistent with exponential decay.
We will confirm this on yet longer time scales in the next section.

The different behaviors of the above two types of observables are based on the differences when operators change or preserve the number parity.
In a very general consideration of a quantum evolution, any observable can be expanded in the eigenstate basis
\begin{equation}
\langle \hat{O}(t) \rangle = \sum_{a,a'} x_a^* O_{a,a'} x_{a'} e^{i (E_a - E_{a'}) t} ~,  
\end{equation} 
where $x_a = \langle a | \psi_{\rm ini} \rangle$, $O_{a,a'} = \langle a| \hat{O} |a' \rangle$, and $|a \rangle$ is an eigenstate with energy $E_a$.
When we consider $\hat{O} = (1/L) \sum_j b_j^\dagger b_{j+1}^\dagger = (1/L) \sum_k c_k^\dagger c_{-k}^\dagger e^{ik}$ (appropriate for calculating $\Upsilon(t)$ in translationally invariant setups), it connects states that differ by precisely two quasiparticles with opposite momenta.
The energy differences can only be $\epsilon_k + \epsilon_{-k} = 2\epsilon_k$.
These are precisely the frequencies that appear in Eq.~(\ref{bbeq}) (see also derivation in App.~\ref{app:BB}).
Note that while the number of states is exponentially large in system size, the number of different frequencies that appear here is only linear in system size, and this is ultimately responsible for the slow power-law ``decoherence" in the observable.
On the other hand, when we consider $\hat{O} = (1/L) \sum_j b_j^\dagger$, the mismatch between the sectors with different particle number parity (related to $b_j^\dagger$ not being locally represented in terms of the JW fermions) results in a much larger number of different frequencies $E_a - E_{a'}$ that appear with non-zero matrix elements; we believe that this is responsible for the faster decay than power law---exponential decay in this case.

To put these results in perspective, the difference in relaxation dynamics of operators that are non-local (contain string) or local (no string) in terms of the diagonalizing Jordan-Wigner fermions has been known in the context of quenches in the quantum Ising chain, starting from Ref.~\cite{Rossini2010} and very detailed subsequent works Refs.~\cite{Calabrese2011, Calabrese2012, Calabrese2012a} (for a recent review, see Ref.~\cite{Essler2016}).
Direct analogs of our $\Phi(t)$ and $\Upsilon(t)$ observables are $\langle \sigma_j^x(t) \rangle$ and $\langle \sigma_j^x(t) \sigma_{j+1}^x(t) \rangle$ in the quantum Ising chain $H = \sum_j (-J \sigma_j^x \sigma_{j+1}^x - h \sigma_j^z)$, which were shown to have exponential and power-law $t^{-3/2}$ envelopes respectively.
It has been also anticipated that such difference holds for other models with free-fermion spectrum.
To our knowledge, our work is the first explicit study of the exponential decay of the order parameter in the case of the BEC to hard-core boson quench.
Our results in the present subsection are numerical, while analytical results for this quench are not available because the time-evolved state here does not have Wick's theorem for the JW fermions, as emphasized in Ref.~\cite{Kormos2014}.
We will present (semi)-analytical results on such a quench in the next section by starting with a different initial state which is qualitatively in the same BEC phase but does have Wick's theorem (and will in fact be able to say more about the product BEC states as well).

\begin{figure}
   \includegraphics[width=0.98\columnwidth]{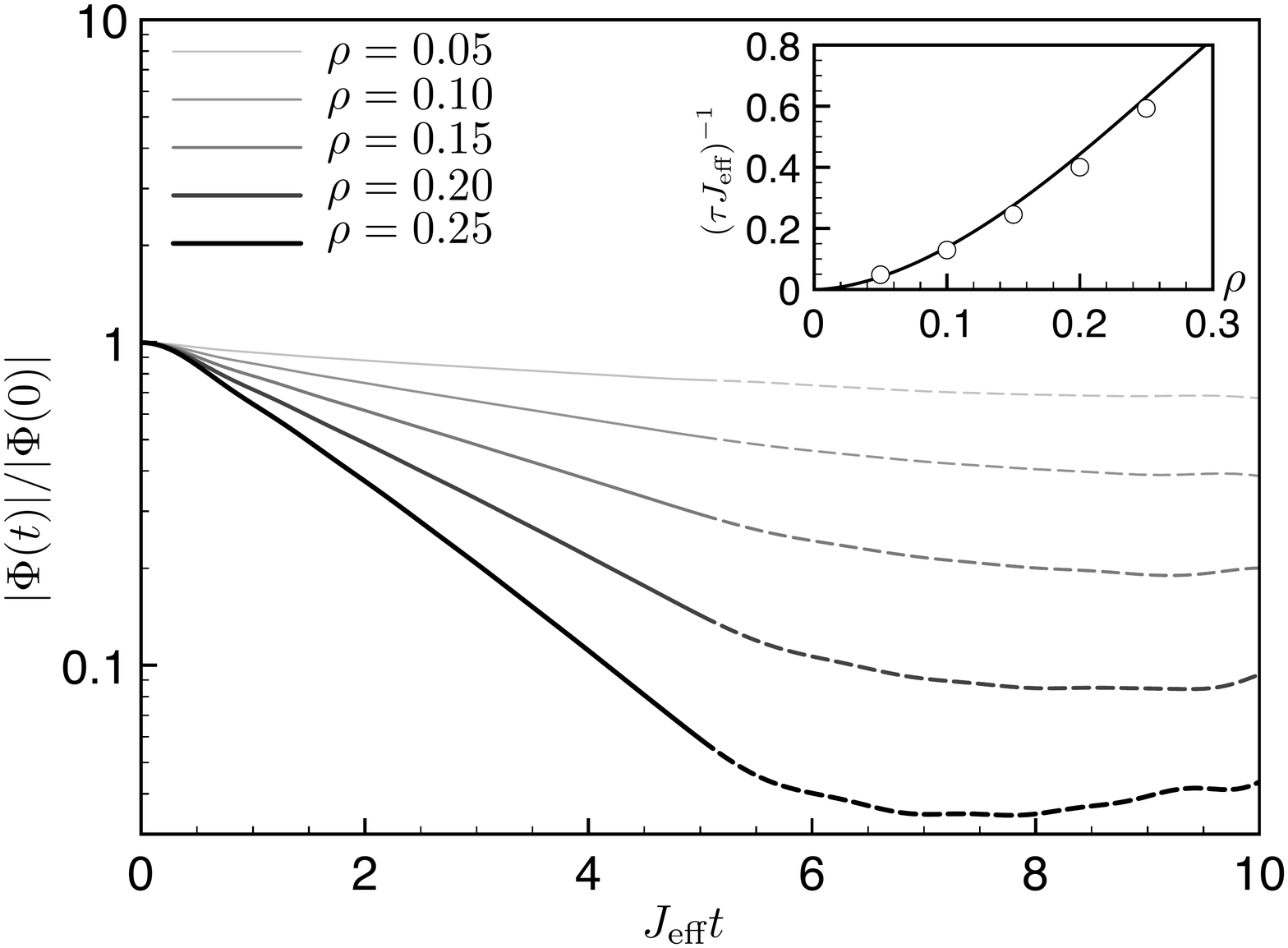}
   \caption{\label{fig:origdensity}
Evolution of the absolute value of the BEC order parameter $|\Phi(t)|$ in the same setting as in Fig.~\ref{bevol} but for different average boson densities $\rho$ in the initial state and showing only the largest ED size $L = 20$ ($\rho = 0.25$ data is the same as in Fig.~\ref{bevol}).
The measurements are normalized by their initial value in order to compare the decay rates.
We clearly see that the decay rate increases as the density increases.
The exponential decay ends at roughly $J_{\text{eff}}t \approx 5$, where the finite-size recurrence effect shows up, indicated by the broken line (see also Fig.~\ref{bevol}).
Inset: density dependence of the inverse lifetime $1/J_{\text{eff}}\tau$.
The circle symbols are obtained from fitting the exponential decay regime $J_{\text{eff}}t \in [1.5, 4.5]$ to a function $A e^{-t/\tau}$, while the solid line is calculated from the conjectured Eq.~(\ref{lifetimerho}).
The inverse lifetime decreases as density decreases, with a $\rho^2 \log(\frac{1}{\rho})$ dependence at small density.
}
   \includegraphics[width=\columnwidth]{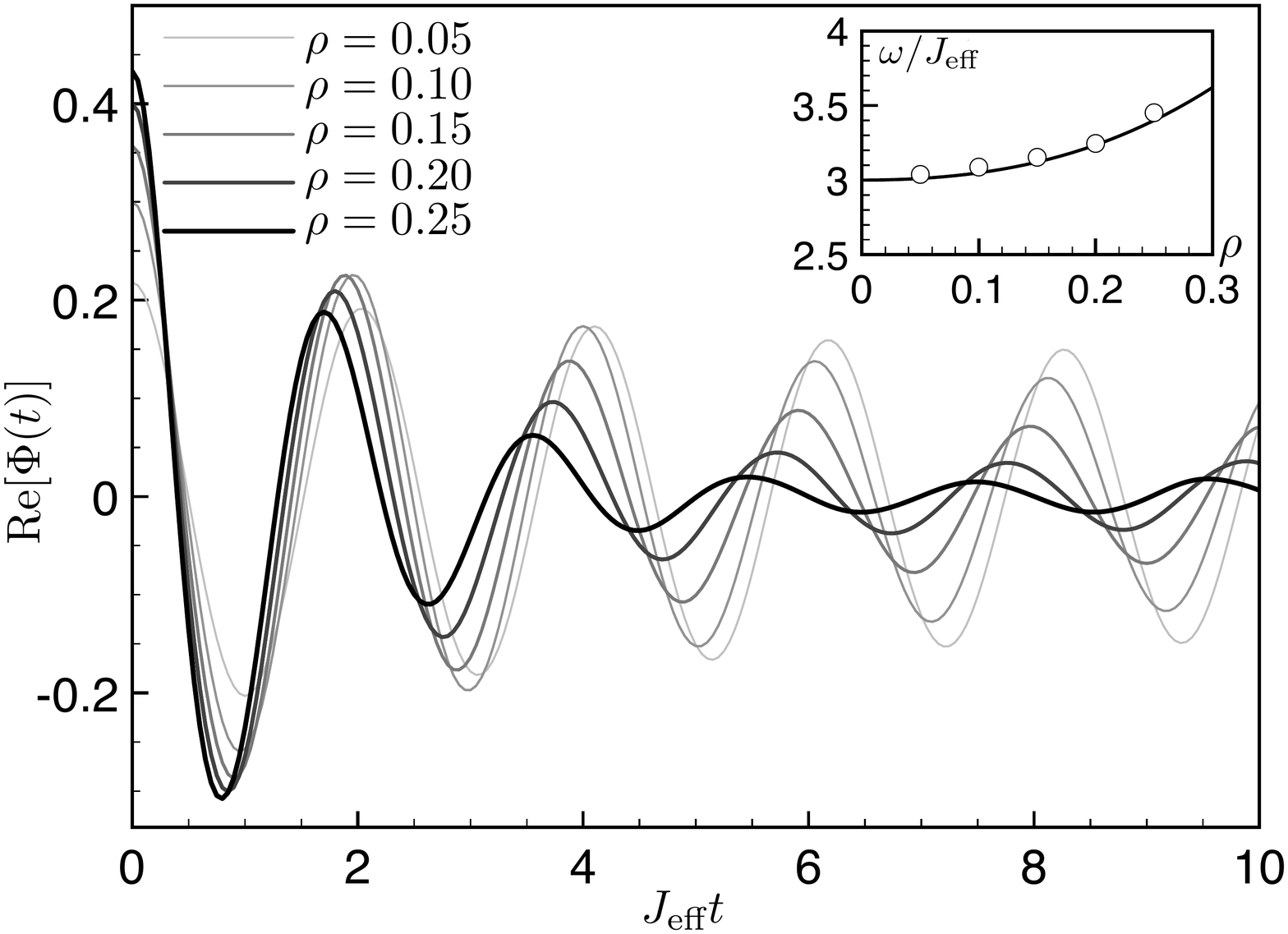}
   \caption{\label{fig:origredensity}
Evolution of the real part of the BEC order parameter ${\rm Re}[\Phi(t)]$ for different average boson densities $\rho$ and showing only the largest ED size $L = 20$; the systems are the same as in Fig.~\ref{fig:origdensity}.
Inset: density dependence of the frequency $\omega$ obtained from fitting the exponential decay regime $J_{\text{eff}}t \in [1.5, 4.5]$ to a function $A e^{-t/\tau} \cos(\omega t - \alpha)$, with $\tau$ determined from Fig.~\ref{fig:origdensity}.
We see that the frequency decreases towards $W_{\rm eff} - 2 J_{\rm eff} = 3$ as the density decreases to zero.
The solid line indicates our conjectured dependence of the oscillation frequency on density, Eq.~(\ref{omegarho}).
}
\end{figure}

Having found exponential decay of the BEC order parameter for a sizable (but otherwise generic) average boson density $\rho = 0.25$ in the initial BEC state, we believe that the same qualitative behavior will persist for all densities.
We will establish this even more firmly on much larger systems and much longer times in the next subsection using a somewhat different realization of the initial BEC state.
Here we would like to study density dependence of the relaxation time moving towards regime of low density, which is of interest in the original spin model. 
Figure~\ref{fig:origdensity} shows $|\Phi(t)|$ evolution for varying average boson density in the initial state.
In each case, we normalized the observable by its initial value in order to get a better comparison.
We can clearly see that the BEC order parameter decays faster with increasing density.
Even though we do not have an exact functional form for $|\Phi(t)|$, we still select the exponential decay regime and fit it with $A e^{-t/\tau}$, where the inverse lifetime $\tau^{-1}$ as a function of density is shown in the inset and vanishes at low density.
A companion Fig.~\ref{fig:origredensity} shows the real part ${\rm Re}[\Phi(t)]$ for the same systems, where we can see that the frequency of oscillations also depends on the density, approaching the $k = 0$ quasiparticle gap $W_{\rm eff} - 2 J_{\rm eff}$ in the limit of low density.

The exponential decay of the order parameter was also obtained in earlier studies of the non-equilibrium dynamics of the magnetization in the quantum Ising model \cite{Sachdev1997, Rossini2010, Calabrese2012} (see also Ref.~\cite{Essler2016} for a recent review).
Despite the differences in details between the hard-core boson and Ising models, the exponential decoherence can be attributed to the non-local nature of the observable when expressed in terms of the Jordan-Wigner fermions, which are the non-interacting quasiparticles in both models.
The origin of the decoherence of the order parameter is the destructive interference coming from contributions from quasiparticles at all momenta.
Ref.~\cite{Calabrese2012} obtained analytical formulas for the decay time and the oscillation frequency in the quantum Ising quench from the ferromagnetic phase to the paramagnetic phase.
These formulas depend only on the mode occupation numbers of the JW fermions in the initial state and not any other details.
We conjecture that the same formulas are valid also for our hard-core boson quench from the BEC state.
We propose the inverse decoherence time (inverse lifetime) as
\begin{equation}
\label{lifetime}
\tau^{-1} = \int_{-\pi}^\pi \frac{dk}{2\pi} \left| \frac{d \epsilon_k}{dk} \right| \log\left| 1 - 2 \langle n_k \rangle \right| ~,
\end{equation}
where $d \epsilon_k/dk$ is the group velocity of the quasiparticle and $\langle n_k \rangle$ is the mode occupation number in the initial state, $\langle n_k \rangle = \langle \psi_{\rm ini} |c_k^\dagger c_k| \psi_{\rm ini} \rangle$.
For the initial state $|\psi_{\rm ini} \rangle = |\psi_{\rm ini, A} \rangle$, calculations similar to those in App.~\ref{app:BB} give in the thermodynamic limit
\begin{equation}
\label{modeoccupationA}
\langle n_k \rangle = \frac{\rho^2 (1+\cos k)}{\rho^2 + (1-\rho)^2 - (1-2\rho) \cos k} ~.
\end{equation}
We can therefore obtain explicit density dependence of the inverse lifetime as
\begin{equation}
\label{lifetimerho}
(J_{\rm eff} \tau)^{-1} = \frac{16}{\pi} ~ \frac{\rho^2 (1-\rho)^2}{\rho^2 + (1-\rho)^2} ~ \frac{\log(1-\rho) - \log(\rho)}{1-2\rho} ~.
\end{equation}
Note that this expression is symmetric under particle-hole transformation sending $\rho \to 1-\rho$, as is expected from simple considerations about this quench.
Importantly for our applications to the original spin model, we find that at low density $(J_{\rm eff} \tau)^{-1} \sim \rho^2 \log(\frac{1}{\rho})$.
Inset in Fig.~\ref{fig:origdensity} compares the inverse lifetime extracted from fits of the time evolution in our ED systems and the conjectured expression Eq.~(\ref{lifetimerho}), denoted by circles and solid line respectively.
The fairly good agreement between the two supports our conjecture.

As for the density dependence of the frequency, based on the quantum Ising study in Ref.~\cite{Calabrese2012}, we can also conjecture that the frequency is given as $\omega = \epsilon_{k_0}$, where $k_0$ is the wave vector such that $1 - 2 \langle n_{k_0} \rangle = 0$.
We can then obtain the frequency as a function of density as 
\begin{equation}
\label{omegarho}
\omega(\rho) = W_{\rm eff} - 2J_{\rm eff} \frac{1 - 2\rho}{\rho^2 + (1-\rho)^2} ~.
\end{equation}
Inset of Fig.~\ref{fig:origredensity} compares the fitted frequencies from the ED study (circles) and the above formula (solid line). 
The close agreement supports our conjecture.

On the other hand, the power-law decay of the pair-boson observable $\Upsilon(t)$ does not depend on the density in the initial state, nor does the oscillation frequency.
This is similar to results in the quantum Ising quench for operators that do not change the Ising quantum number \cite{Rossini2010, Calabrese2012, Essler2016} and
is also another noteworthy point of the differences between the relaxation behaviors of the single- and pair-boson operators.

\subsection{Initial hard-core boson BEC state realized as a topological superconductor of JW fermions}
\label{sec:topoini}
The initial state used in the previous subsection does not have any special properties like the Wick's theorem that we could utilize to reach larger system sizes when calculating the evolution of the BEC order parameter.
In this subsection, we will consider a different initial state which is qualitatively in the same BEC phase but for which Wick's theorem is valid, therefore enabling calculations for much larger system sizes.

Specifically, consider the following hard-core boson Hamiltonian,
\begin{eqnarray}
H_{\rm ini} &=& -J_0 \sum_{j=1}^L (b_j^\dagger b_{j+1} + \Hc) \nonumber \\
&-& \Delta_0 \sum_{j=1}^L (b_j^\dagger b_{j+1}^\dagger + \Hc) - \mu_0 \sum_{j=1}^L b_j^\dagger b_j ~,
\label{Hb4toposc}
\end{eqnarray}
with periodic boundary conditions, $b_{j+L} \equiv b_j$.
This Hamiltonian preserves particle number parity, with the corresponding ground states in the even and odd parity sectors $|\psi_{\rm g.s., even} \rangle$ and $|\psi_{\rm g.s., odd} \rangle$.
We will argue below that as long as $|\mu_0| < 2 |J_0|$, these ground states have long-range order in the single-boson observable, i.e., $\lim_{|j-j'| \to \infty} \langle b_j^\dagger b_{j'} \rangle \neq 0$.
Schematically, we can indicate this long-range order by writing $\langle b_j^\dagger \rangle \neq 0$.
In particular, it is actually sensible to consider a superposition of $|\psi_{\rm g.s., even} \rangle$ and $|\psi_{\rm g.s., odd} \rangle$ and view it as a BEC of bosons, which contains states with arbitrary particle numbers.
For example, we can take $|\psi_{\rm ini} \rangle = (1/\sqrt{2}) (|\psi_{\rm g.s., even} \rangle + |\psi_{\rm g.s., odd} \rangle)$, which for $J_0, \Delta_0 > 0$ and appropriate choices of the phases of $|\psi_{\rm g.s., even/odd} \rangle$ will have positive amplitudes on all states in the boson number basis, similarly to the state Eq.~(\ref{hcbBEC}) with real and positive parameters $\alpha$ and $\beta$.
However, as will become clear, the details of the superposition are not important.

The above Hamiltonian can be also exactly solved by the Jordan-Wigner transformation Eq.~(\ref{JWtransform}), which gives
\begin{eqnarray}
H_{\rm ini} &=& -J_0 \sum_{j=1}^{L-1} (c_j^\dagger c_{j+1} + \Hc) - \Delta_0 \sum_{j=1}^{L-1} (c_j^\dagger c_{j+1}^\dagger + \Hc) \nonumber \\
&-& J_0 (-1)^{N_{\rm tot} + 1} (c_L^\dagger c_1 + \Hc) \nonumber \\
&-& \Delta_0 (-1)^{N_{\rm tot} + 1} (c_L^\dagger c_1^\dagger + \Hc) - \mu_0 \sum_{j=1}^L c_j^\dagger c_j ~.
\end{eqnarray}
The fermions effectively have antiperiodic boundary conditions in the even number-parity sector and periodic boundary conditions in the odd number-parity sector.
After Fourier transformation with lattice momenta $k = \frac{2\pi}{L}(m + \frac{1}{2})$, $m = 0, 1, \dots, L-1$ in the even number-parity sector and $k = \frac{2\pi}{L}m$ in the odd number-parity sector, we further apply Bogoliubov transformation to diagonalize the above Hamiltonian.
The Bogoliubov quasiparticles are given as $\gamma_k = u_k^* c_k + v_k^* c_{-k}^\dagger$, with $u_k = \cos(\theta_k/2)$, $v_k = \sin(\theta_k/2)$. 
The parameter $\theta_k$ is determined by $\tan(\theta_k) = \frac{-2\Delta_0 \sin(k)}{2J_0\cos(k) + \mu_0}$. 
We can readily construct the vacuum of the Bogoliubov quasiparticles in both parity subspaces; e.g., in the even-parity sector we have 
$|\psi_{\rm g.s., even} \rangle = |{\rm vac}_{\gamma, {\rm even}} \rangle = \prod_{k > 0} \left( u_k^* - v_k^* c_k^\dagger c_{-k}^\dagger \right)|{\rm vac}_c  \rangle$, 
where $|{\rm vac}_c \rangle$ is the vacuum for the $c$ fermions and $k$ are chosen appropriately for this number-parity.

The dynamics is governed by the hard-core boson hopping Hamiltonian Eq.~(\ref{Hhcb}).
We are interested in the time evolution of the BEC order parameter $\Phi(t)$ and will calculate its real part,
$2 {\rm Re}[\langle b_j^\dagger(t) \rangle] = \langle b_j^\dagger(t) + b_j(t) \rangle \equiv \langle \Sigma_j(t)\rangle$.
The site index $j$ can be arbitrary since the state remains translationally invariant during the evolution. 
Again, exactly solving the time evolution of $\Sigma_j$ is very difficult due to the mismatch between the JW fermion boundary conditions in the even and odd sectors.
In order to remedy this obstacle, we adopt the factorization trick of McCoy\etal~\cite{McCoy1971}.
Instead of considering $\langle \Sigma_j(t) \rangle$ directly, we consider
\begin{equation}
\langle \Sigma_j(t) \Sigma_{j+\ell}(t) \rangle \approx \langle \Sigma_j(t) \rangle \langle \Sigma_{j+\ell}(t) \rangle
\label{factortrick}
\end{equation} 
for separations $\ell \gg vt$, where $v$ is some characteristic velocity for the spreading of quantum correlations. 
In this limit, we expect that the above approximation is very accurate based on reasoning similar to Lieb-Robinson bound \cite{Hastings2004}, although we have not tried to prove this rigorously. 

Since $\Sigma_j \Sigma_{j+\ell}$ does not mix the even and odd sectors, the above trick enables us to deal separately with the two sectors.
Furthermore, it is sufficient to consider the even sector only and calculate the quantity 
\begin{equation}
R(\ell, t) \equiv \langle \psi_{\rm ini,B} | \Sigma_j(t) \Sigma_{j+\ell}(t) |\psi_{\rm ini,B} \rangle ~,
\end{equation}
where we choose the initial state as $|\psi_{\rm ini,B} \rangle=|\psi_{\rm g.s.,even} \rangle$, since we expect the contribution from $|\psi_{\rm g.s.,odd} \rangle$ will essentially be identical in the thermodynamic limit \cite{Calabrese2012}.

In the fermionic representation,
\begin{eqnarray}
R(\ell, t) =
\langle\psi_{\rm ini,B}|\left( \prod_{j'=j}^{j+\ell-1} B_{j'}(t) A_{j'+1}(t) \right) | \psi_{\rm ini,B} \rangle ~,~~
\label{BABA}
\end{eqnarray}
where we defined Majorana fermions $A_j = c_j^\dagger + c_j$ and $B_j = c_j^\dagger - c_j$.
It is easy to perform calculations in the Schrodinger picture of the time evolution.
Thus, 
\begin{eqnarray}
|\psi_{\rm ini,B}(t) \rangle &=& e^{-iHt} \prod_{k>0} (u_k^* - v_k^* c_k^\dagger c_{-k}^\dagger) e^{iHt} e^{-iHt} |{\rm vac}_c \rangle \nonumber \\
&=& \prod_{k>0} (u_k^* - v_k^* c_k^\dagger c_{-k}^\dagger e^{-i 2\epsilon_k t}) |{\rm vac}_c \rangle\nonumber \\
&=& \prod_{k>0} [u_k^* - v_k^*(t) c_k^\dagger c_{-k}^\dagger] |{\rm vac}_c \rangle ~, 
\end{eqnarray} 
where the dynamics can be considered as an evolution of the coherence factor $v_k(t) \equiv v_k e^{2 i \epsilon_k t}$.

Since the above state can be viewed as a BCS ground state of a Hamiltonian with the corresponding coherence factors at every instant, the Wick's theorem holds for $|\psi_{\rm ini,B}(t) \rangle$ at every time $t$. 
In order to apply the Wick's theorem, we need to evaluate the following two-operator correlation functions:
\begin{eqnarray*}
\langle A_m A_n \rangle &=& \frac{1}{L} \sum_k \left[1 - \sin\theta_k \sin(2\epsilon_k t) \right] e^{i k (m-n)} ~, \\
\langle B_m B_n \rangle &=& \frac{1}{L} \sum_k \left[-1 - \sin\theta_k \sin(2\epsilon_k t) \right] e^{i k (m-n)} ~, \\
\langle A_m B_n \rangle &=& \frac{1}{L} \sum_k \left[\cos\theta_k + i \sin\theta_k \cos(2\epsilon_k t) \right] e^{i k (m-n)} ~, \\
\langle B_m A_n \rangle &=& \frac{1}{L} \sum_k \left[-\cos\theta_k + i \sin\theta_k \cos(2\epsilon_k t) \right] e^{i k (m-n)} ~.
\end{eqnarray*}
For conciseness, we define Toeplitz matrices with elements 
$[{\tt AA}]_{m, n} = \langle A_m A_n \rangle$ when $m \neq n$ and
$[{\tt AA}]_{m, n} = 0$ when $m = n$;
$[{\tt BB}]_{m, n} = \langle B_m B_n \rangle$ when $m \neq n$ and
$[{\tt BB}]_{m, n} = 0$ when $m = n$;
$[{\tt BA}]_{m, n} = \langle B_m A_{n+1} \rangle$; and 
$[{\tt AB}]_{m, n} = \langle A_{m+1} B_n \rangle$.
$[{\tt AA}]$ and $[{\tt BB}]$ are antisymmetric matrices while $[{\tt AB}] = -[{\tt BA}]^T$.
We then define a $2\ell \times 2\ell$ matrix $M$ as a block Toeplitz matrix with elements
\begin{equation}
\begin{pmatrix}
M_{2m-1, 2n-1} & M_{2m-1, 2n} \\
M_{2m, 2n-1} & M_{2m, 2n}
\end{pmatrix}
= \begin{pmatrix}
[{\tt BB}]_{m, n} & [{\tt BA}]_{m, n} \\
[{\tt AB}]_{m, n} & [{\tt AA}]_{m, n}
\end{pmatrix} ~,
\end{equation}
where $m, n = 1, \dots, \ell$. 
Note that the matrix $M$ is antisymmetric.
Applying Wick's theorem to Eq.~(\ref{BABA}) with $j=1$, we then have $R(\ell, t) = \text{Pf}(M)$, the Pfaffian of the above matrix $M$.  

Before discussing the time evolution, let us consider properties of the initial state encoded in $R(\ell, t=0)$, which is a specific boson-boson correlation function in the initial state $|\psi_{\rm ini,B} \rangle$.
This correlation function exhibits two different behaviors depending on the parameters of $H_{\rm ini}$.
At $t=0$, matrices $[{\tt AA}]$ and $[{\tt BB}]$ are zero, and by rearranging the columns and rows of the matrix $M$, we obtain
\begin{equation*}
R(\ell, t=0) = (-1)^{\ell(\ell-1)/2} \text{Pf}
\begin{pmatrix}
0 & [{\tt BA}] \\
-[{\tt BA}]^{T} & 0
\end{pmatrix} 
= \det[{\tt BA}] ~,
\end{equation*}
where matrix $[{\tt BA}]$ is evaluated at $t=0$.
Thus, $R(\ell, t=0)$ is equal to the determinant of the Toeplitz matrix $[{\tt BA}]$.
The asymptotic behavior at large $\ell$ is given by Szeg\"o's theorem \cite{McCoy1973},
\begin{equation}
\lim_{\ell \to \infty} R(\ell, t=0) \sim Ce^{\lambda_0 \ell}  ~,
\end{equation} 
where $C$ is a constant and $\lambda_0=\int_{-\pi}^{\pi}\frac{dk}{2\pi}\log(e^{-i \theta_k} e^{-i k})$. 
When $\lambda_0 \neq 0$, which occurs for $|\mu_0| > 2|J_0|$, we have exponential decay of the correlation function.
On the other hand, when $\lambda_0 = 0$, which occurs for $|\mu_0| < 2|J_0|$, the correlation function approaches a non-zero constant, signaling a long-range order in the boson BEC order parameter.
Note that in terms of the Jordan-Wigner fermions, conditions $|\mu_0| > 2|J_0|$ and $|\mu_0| < 2|J_0|$ correspond respectively to the trivial and topological superconductor phase in the one-dimensional spinless superconductor \cite{Kitaev2001, Motrunich2001} (i.e., strong-coupling and weak-coupling superconducting phases in the sense of Read and Green, Ref.~\cite{Read2000}).
Thus, we have analytically proven an earlier numerical finding in Ref.~\cite{Motrunich2012} that the topological phase of JW fermions corresponds to the single-boson long-range order, while the trivial phase corresponds to short-range order; both phases clearly have long-range order in the pair-boson correlator.
We then choose our initial state to be in the regime of the weak-coupling (topological) phase of the JW fermions, which hence has non-vanishing long-range order (BEC) in terms of the original bosons.

\begin{figure}
    \includegraphics[width=\columnwidth]{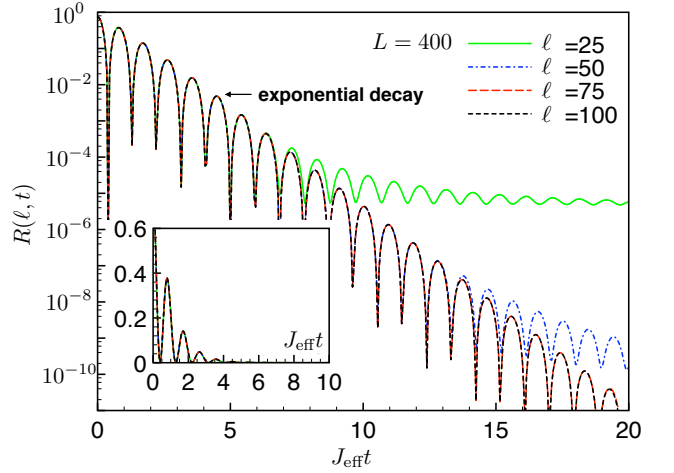}
    \caption{\label{fig:topobevol} (color online)
Numerical results for $R(\ell, t)$ obtained using the Pfaffian method; the system size is $L = 400$, and we consider separations $\ell = 25, 50, 75$, and $100$.
The initial state is the ground state of Eq.~(\ref{Hb4toposc}) with parameters $\Delta_0 = 0.60$ and $\mu_0=-1.60$, corresponding to particle density $\rho = 0.25$.
The parameters of the dynamical Hamiltonian are chosen as $J_{\rm eff} = 1$ and $W_{\rm eff} = 5$.
We are primarily interested in the regime $t \ll \ell/v$, where $v$ is some information-spreading velocity.
In this regime, the observable exhibits exponential decay. 
The time interval over which we see the exponential decay increases as the separation $\ell$ increases.}
\end{figure}

\begin{figure}
   \includegraphics[width=\columnwidth]{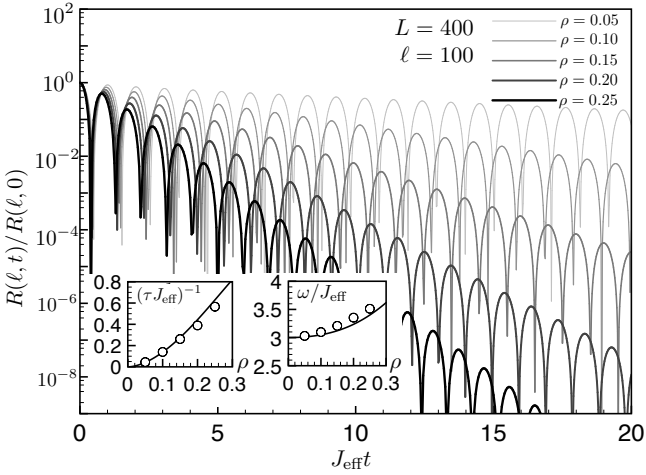}
   \caption{\label{fig:topodensity} 
$R(\ell, t)$ as in Fig.~\ref{fig:topobevol} but for different average densities $\rho$ obtained by tuning $\mu_0$ and $\Delta_0$ according to Eq.~(\ref{muurho}) and Eq.~(\ref{Deltarho}) respectively.
We show only the largest separation $\ell = 100$; the full system size is $L = 400$ and is sufficiently large to reflect the thermodynamic limit in $L$.
The values are normalized by the initial value in order to bring out the decay rate, which clearly increases as the density increases.
Insets: inverse relaxation time $\tau^{-1}$ of the exponential decay and the oscillation frequency $\omega$ as a function of density, obtained from fitting the data in the main panel to form $A e^{-2t/\tau} \cos^2(\omega t - \alpha) + C$.
The circle symbols denote the fitted values, while the solid lines denote the conjectured forms Eq.~(\ref{lifetimerho}) and Eq.~(\ref{omegarho}) for $\tau^{-1}$ and $\omega$ respectively (see text for details).}
\end{figure}

After specifying the suitable initial state, we can now discuss the dynamics. Figure~\ref{fig:topobevol} shows $R(\ell, t)$ calculated for various separations $\ell$ in a system with total length $L = 400$.
We are interested in the regime where $t \ll \ell/v$, where $v$ is some characteristic velocity for the information spreading. 
We unambiguously see that $R(\ell, t)$ shows an exponential decay over some time interval that increases with increasing separation $\ell$.
This behavior corresponds to the exponential decay of $\langle \Sigma_j(t) \rangle$ with time, as claimed earlier for the BEC order.

On the other hand, we can also consider $\langle \Sigma_j(t) \Sigma_{j+1}(t) \rangle$, which is similar to the pair-boson observable we considered earlier that does not change the particle number parity.
In this case,
\begin{eqnarray}
&& \langle \Sigma_j(t) \Sigma_{j+1}(t) \rangle = [{\tt BA}]_{1,1} = \langle B_1 A_2 \rangle \nonumber \\
&& = \frac{1}{L} \sum_k \left[-\cos\theta_k + i \sin\theta_k \cos(2\epsilon_k t) \right] e^{-ik} ~.
\end{eqnarray}
At long time, this approaches a constant value given by $\langle b^\dagger_j(t) b_{j+1}(t) \rangle + {\rm c.c.}$.
The time-dependent part comes from $\langle b^\dagger_j(t) b^\dagger_{j+1}(t) \rangle + {\rm c.c.}$; upon using the steepest descent analysis, we find oscillations at frequencies $2\epsilon_{k=0/\pi}$ with a power law envelope $t^{-3/2}$, similar to results in the previous subsection for the product BEC initial state.
We expect similar behaviors for any fixed $\ell$ at long times $t \gg \ell/v$:
Indeed, we expect $\lim_{t \to \infty} \langle b^\dagger_j(t) b_{j+\ell}(t) \rangle = \langle b^\dagger_j b_{j+\ell} \rangle_{\rm therm.} \equiv C_{\rm therm.}(\ell) \neq 0$, where $C_{\rm therm.}(\ell)$ decays exponentially with $\ell$.
On the other hand, we expect $\lim_{t \to \infty} \langle b^\dagger_j(t) b^\dagger_{j+\ell}(t) \rangle = 0$, where the approach to zero has a power-law envelope $\sim t^{-3/2}$.
These predictions are based on our expectations that at long times local observables can be described using a generalized Gibbs ensemble which is diagonal in the particle number, and that for large $t \gg \ell/v$ the physics of $b^\dagger_j(t) b^\dagger_{j+\ell}(t)$ is that of a local pair-boson creation operator.
   
We thus see significant care needed when using $R(\ell, t)$ to extract the behavior of the BEC order parameter $\langle b_j^\dagger(t) \rangle$ using Eq.~(\ref{factortrick}) which holds only for $t \ll \ell/v$.
In Fig.~\ref{fig:topobevol} we chose to show $\langle \Sigma_j(t) \Sigma_{j+\ell}(t) \rangle$ with $\Sigma_j = b^\dagger_j + b_j$ so that the regime where the sites $j$ and $j+\ell$ start to ``feel each other" is manifest by approaching a constant due to $\langle b^\dagger_j(t) b_{j+\ell}(t) \rangle + {\rm c.c.}$ pieces as discussed above (if we only had $\langle b^\dagger_j(t) b^\dagger_{j+\ell}(t) \rangle + {\rm c.c.}$ pieces, this time scale would manifest as a crossover from the exponential to $t^{-3/2}$ decay and would be more difficult to detect).

In order to study the density dependence of the decoherence time of the long-range order and compare with the results in the previous subsection, it is tempting to tune the parameters of $|\psi_{\rm ini, B} \rangle$, $\mu_0$ and $\Delta_0$, such that the density and the energy density are equal to those in $|\psi_{\rm ini, A} \rangle$.
In fact, when trying to achieve this, we found that it is possible to make the JW fermion mode distribution $\langle n_k \rangle$ identical in the two initial states in the thermodynamic limit!
The mode occupation number in $|\psi_{\rm ini, B} \rangle$ is given as $\langle n_k \rangle = |v_k|^2$, or,  
\begin{equation}
\langle n_k \rangle = \frac{1}{2} \left(1 - \frac{-2J_0 \cos(k) - \mu_0}{\sqrt{(2J_0 \cos(k) + \mu_0)^2 + [2\Delta\sin(k)]^2}} \right) ~.
\end{equation}
One can easily verify that if we take
\begin{equation}
\label{muurho}
\mu_0 = -2\sqrt{J_0^2 - \Delta_0^2}
\end{equation}
and
\begin{equation}
\label{Deltarho}
\Delta_0 = J_0 \frac{2 \rho (1-\rho)}{\rho^2 + (1-\rho)^2} ~,
\end{equation}
the mode occupation number will become identical to Eq.~(\ref{modeoccupationA}).
Here we assumed $\rho < 0.5$, while for $\rho > 0.5$ we need to take the opposite sign for $\mu_0$ in Eq.~(\ref{muurho}).
In retrospect, by examining the JW fermion pair-function in real-space for the topological superconductor with the condition Eq.~(\ref{muurho}), we can show that the many-body wavefunction $|\psi_{\rm ini, B} \rangle$ in the boson representation exactly coincides with that of the BEC state Eq.~(\ref{hcbBEC}) when {\it projected to any sector with fixed particle number}. 
The relative weights on the different number sectors do not coincide for $|\psi_{\rm ini, A} \rangle$ and $|\psi_{\rm ini, B} \rangle$, but this is not important in the thermodynamic limit.

Figure~\ref{fig:topodensity} shows $R(\ell, t)$ with $\ell = 100$ and system size $L = 400$ for several different densities $\rho = 0.05$ to $\rho = 0.25$, with $\mu_0$ and $\Delta_0$ chosen according to Eq.~(\ref{muurho}) and Eq.~(\ref{Deltarho}).
We use the formula $Ae^{-2t/\tau}\cos^2(\omega t - \alpha)+C$ to fit our numerical data hence obtaining the fitted lifetime $\tau$ and frequency $\omega$.
The left inset compares the fitted inverse lifetime $\tau^{-1}$ denoted by circles and the conjectured inverse lifetime based on Eq.~(\ref{lifetimerho}) denoted by solid line.
The agreement between the two is fairly good and therefore supports our conjecture.
The right inset compares the fitted frequency (denoted as circles) and the conjectured frequency (solid line) based on Eq.~(\ref{omegarho}).
The oscillation frequency also agrees with the conjecture quite well. 

We hence see that the new choice of the initial state $|\psi_{\rm ini,B} \rangle$, equipped with Wick's theorem, enables us to calculate the evolution for much larger system size (essentially in the thermodynamic limit). 
We therefore further confirm the exponential decay of the BEC order parameter and provide strong numerical evidence for our conjecture of the density dependence of the inverse lifetime and frequency.

\begin{figure}
    \includegraphics[width=\columnwidth]{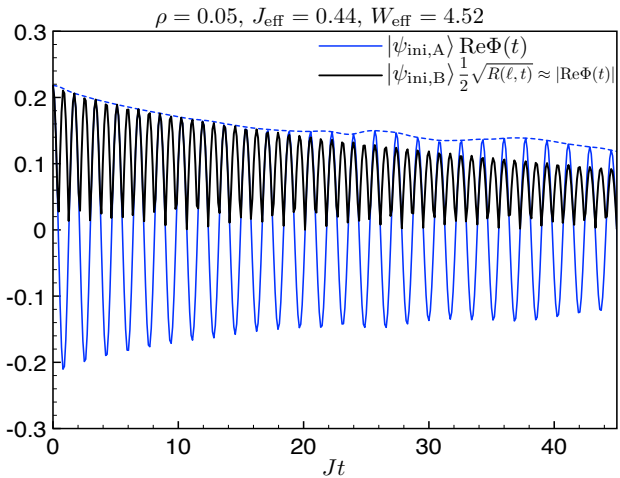}
    \caption{\label{fig:spinRelEvol} (color online)
Evolution of $\text{Re}[\Phi(t)]$ for the initial state $|\psi_\text{ini, A} \rangle$ and system size $L = 20$ and $|\text{Re}[\Phi(t)]|$ calculated from $\frac{1}{2}\sqrt{R(\ell, t)}$ for the initial state $|\psi_\text{ini, B} \rangle$ with separation $\ell = 100$ and system size $L=400$, both with density $\rho = 0.05$.
The dynamical Hamiltonian has $J_\text{eff} = 0.44$ and $W_\text{eff} = 4.52$, chosen to reproduce the spin-flip quasiparticle dispersion and gap $W_{\rm eff} - 2J_{\rm eff} = \Delta E = 3.64$ in the original spin problem.
The relaxation times obtained by fitting the exponential decay regime as in Fig.~\ref{fig:origredensity} and Fig.~\ref{fig:topodensity} are $\tau_A \approx 46.5$ and $\tau_B \approx 51.0$ respectively.
This means that at time $t = 18$---the longest time simulated in the original spin problem---the amplitude will be roughly $0.7$ of the initial value.}
\end{figure}

\subsection{Application to the original spin problem}
\label{sec:cfspin}

Returning to the original spin problem, in order to make more quantitative comparisons, Fig.~\ref{fig:spinRelEvol} shows results for the initial states $|\psi_{\rm ini, A} \rangle$ and $|\psi_{\rm ini, B} \rangle$ evolving under the hard-core boson Hamiltonian with parameters $J_{\rm eff} = 0.44$, $W_{\rm eff} = 4.52$, and density $\rho = 0.05$.
These are chosen to be close to the parameters of the spin-flip quasiparticles of the original spin problem, cf.\ Sec.~\ref{truncatedSW}.
By fitting the exponential decay regime, we obtain the decoherence time $\tau_A \approx 46.5$ and $\tau_B \approx 51.0$ respectively, which are close to an estimate from Eq.~(\ref{lifetimerho}) that gives $\tau \approx 54.7$ for these $J_{\rm eff}$ and $\rho$.
The long lifetime is due to the low density of the quasiparticles (due to low energy density in the initial state), and is the primary reason of the apparent persistent oscillation in the original spin problem.
We see that up to time $t = 18$ simulated in the original spin problem in Ref.~\cite{Banuls2011}, the amplitude of interest decays to roughly $0.7$ of the initial value.
The exponential decay is not easily seen in this time range, while it becomes more clear when one goes to longer times. 
The $L = 20$ ED results in Fig.~\ref{fig:spinRelEvol} again show recurrence phenomenon starting from about $t \sim 15-20$, while the Pfaffian calculation results represent the thermodynamic limit.

We remark that Fig.~\ref{fig:EDevolution} hardly shows any decay while Fig.~\ref{fig:spinRelEvol} shows some gradual decay already for $t \leq 18$.
There could be several reasons for this.
The decrease from the initial value in Fig.~\ref{fig:EDevolution} is actually significant, and it could be that the first few oscillations happened to experience stronger decrease due to some microscopics, which then masked the more systematic decay expected at long times.
In this respect we remind that the hard-core boson model studied in this section neglected all interactions among the quasiparticles of the spin model other than their hard-core exclusion, see Eq.~(\ref{Hprime}) and discussion at the end of Sec.~\ref{subsec:qppic} (and we also remind that the BEC is only an approximation to the initial state of the quasiparticles).
While we would expect that additional interactions generically help the thermalization, we do not know the actual quantitative effect which can depend on details and requires more studies.
We also mention that in the truncated Schrieffer-Wolff approach the observables also obtain components on the pair-boson-type operators and hence will have additional power-law-decaying contributions in the hard-core boson model (which should eventually also decay exponentially once integrability-breaking interactions are included).
All such unaccounted parts could have enough effect up to $t = 18$ to make the oscillation appear more persistent, while they eventually decay at longer times.
We therefore propose that if one can simulate the original spin problem to somewhat longer time, one will eventually observe the decay of the oscillations.

In fact, the Supplemental Material of Ref.~\cite{Banuls2011} also showed a study for parameter $g = -1.5$, where the observables showed visible decays, which we believe can be understood as due to the larger particle density and hence shorter decoherence time.
Our ED calculations give energy density in the $|Z\!+ \rangle$ state over the ground state as $(\langle Z\!+ |H| Z\!+ \rangle - E_0)/L = 0.4581$ and the quasiparticle gap as $E_1 - E_0 = 3.2041$, so the quasiparticle density in the ground state is roughly $\rho \approx 0.143$.
For this density, Eq.~(\ref{lifetime}) gives $J_{\rm eff} \tau \approx 3.935$.
From our ED data, we can also extract the quasiparticle hopping amplitude $J_{\rm eff} \approx 0.656$.
This gives us $\tau \approx 6$, which was already accessible in the infinite-MPS study in Ref.~\cite{Banuls2011}, and our estimates of the oscillation frequency and decay time are in rough agreement with this.

We note another lesson learned from the detailed study of the relaxation of the order parameter in the hard-core boson model.
While our original stipulation was that the oscillation frequency is set by the quasiparticle gap, we now see that it is actually a function of the quasiparticle density, see Eq.~(\ref{omegarho}).
We should indeed expect this generically as the quasiparticle energy is strictly defined only in the limit of vanishing energy density, while here the initial state has a finite energy density (e.g., quasiparticle energies can get renormalized by their residual interactions, etc.).
In this respect, the single-boson observable shows more generic behavior even in the integrable hard-core boson model, in contrast to the pair-boson observable whose oscillation frequencies are independent of the density.
In any case, the original spin model at $g = -1.05$ is at sufficiently low energy density that the oscillation frequency is very close to the quasiparticle gap, and we did not worry about differentiating between these in the previous sections.

\begin{figure}
    \includegraphics[width=0.95\columnwidth]{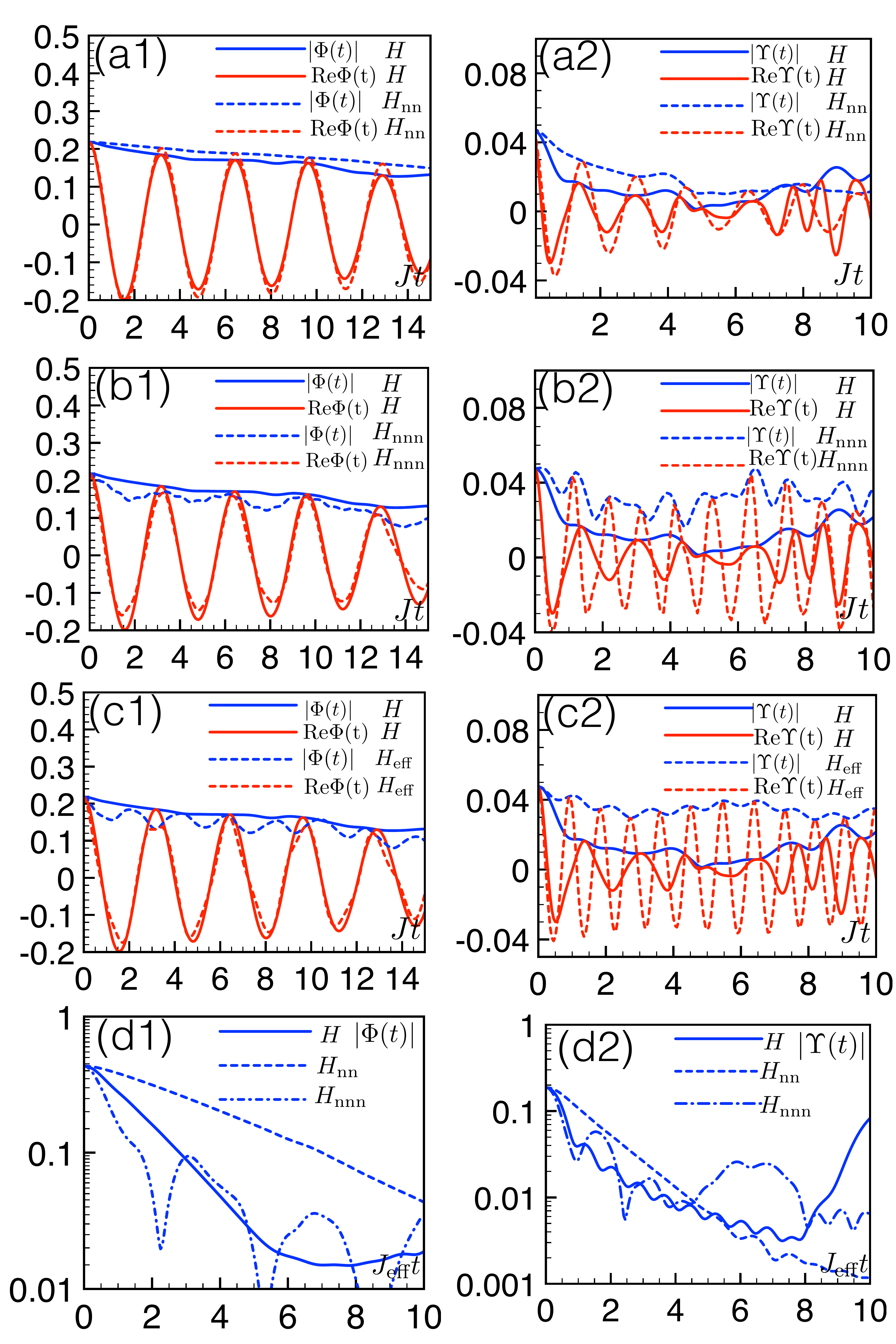}
 \caption{\label{fig:HCBvsTrunc} (color online) 
Comparison of the dynamics under the simplified hard-core boson Hamiltonian $H$ in Eq.~(\ref{Hhcb}) and under various Hamiltonians with different additional interactions. The initial state is $|\psi_{\rm{ini,A}}\rangle$.
(a)-(c) BEC order parameter (left panels) and pair boson correlation function (right panels) for density $\rho=0.05$ and system size $L=14$.
The parameters of $H_{\rm{eff}}$ are obtained from second-order SW for the original spin model, and similar parameters are also used in $H$, $H_{\rm{nn}}$, $H_{\rm{nnn}}$: $J_{\rm{eff}}=0.88$, $W_{\rm{eff}}=3.68$, $W_{\rm{nn}}=-0.47$, and $W_{\rm{nnn}}=2.35$ (in units of $J$). 
The additional interactions basically introduce a small amplitude oscillation in $\Phi(t)$. 
For the pair boson correlation function, the decay behavior and the oscillation frequencies are affected more strongly (see text for details).
(d) BEC order parameter $\Phi(t)$ and pair-boson correlation function $\Upsilon(t)$ for density $\rho=0.25$, system size $L=20$ and parameters $J_{\rm{eff}}=1.0$, $W_{\rm{eff}}=1.0$, $W_{\rm{nn}}=-0.5$, and $W_{\rm{nnn}}=1.5$ (the results for $H$ are identical as in Fig.~\ref{bevol} and ~\ref{bbevol}). 
The additional interaction of the JW fermions in $H_{\rm{nn}}$ changes the decay in $\Upsilon(t)$ to be faster than the previous power law.}
\end{figure}

While the above discussions are based on the hard-core boson hopping Hamiltonian, $H$ in Eq.~(\ref{Hhcb}), we now briefly consider effects of adding more generic interactions.
To understand these, we calculated the BEC order parameter $\Phi(t)$ and the pair-boson correlation function $\Upsilon(t)$ evolving under (a) $H_{\rm{nn}}=H+W_{\rm{nn}}\sum_{j}n_{j}n_{j+1}$; (b) $H_{\rm{nnn}}=H_{\rm{nn}}+W_{\rm{nnn}}\sum_{j}n_{j}n_{j+2}$; and (c) $H_{\rm{eff}}$ from Eq.~(\ref{Hprime}) upon dropping $H_{\rm{other}}$. In each case, the initial state is $|\psi_{\rm{ini,A}}\rangle$.

Plots Fig.~\ref{fig:HCBvsTrunc}~(a)-(c) show the corresponding $\Phi(t)$ and $\Upsilon(t)$ results for system size $L=14$ and particle density $\rho=0.05$, which is close to the original spin model. 
On the other hand, plots in Fig.~\ref{fig:HCBvsTrunc}~(d) shows $\Phi(t)$ and $\Upsilon(t)$ for $\rho=0.25$ and system size $L=20$; this is far from the original spin model but allows us to reduce effects due to low particle density and to focus instead on qualitative effects of interactions.
We find that the additional interaction terms do not change the qualitative exponential decay behavior of $\Phi(t)$.
This is less clear for low density, plots (a1)-(c1), but in these cases, the effect of interactions are also quantitatively small, and we believe the decays are still exponential at long time.
On the other hand, the behavior of the pair-boson correlation function $\Upsilon(t)$ changes qualitatively. 
The power-law decay of $\Upsilon(t)$ in the hard-core boson hopping model is a consequence of the exact solution in terms of the non-interacting JW fermions.
From Fig.~\ref{fig:HCBvsTrunc}~(d2), it appears that as we turn on the interaction of the fermions, the power-law decay behavior is destroyed. 
The addtional interaction in $H_{\rm{nn}}$ turns $\Upsilon(t)$ into a faster decay; under $H_{\rm{nnn}}$, the behavior of $\Upsilon(t)$ is not clear and needs more study. 
While we would expect exponential decay for generic interactions, this appears to be also true for $H_{\rm{nn}}$ which is still integrable; we speculate this is because the true excitations are no longer JW fermions.
 
Finally, we note that for $H_{\rm{eff}}$, the behavior of $\Upsilon(t)$ is somewhat special. 
The reason is that localized bound states of two spin-flip quasiparticles (i.e., two consecutive spin-downs in the background of spin-ups) happen to be exact eigenstates of $H_{\rm{eff}}$, since the correlated hopping terms just annihilate these states. 
Therefore, these bound states are immobile in $H_{\rm{eff}}$, which can be detected by $b^{\dagger}_{j}b^{\dagger}_{j+1}$. 
As a result, the pair-boson correlation function oscillates without decay, which is an artifact of this second-order truncated-SW Hamiltonian.
However, if we introduce dynamics for these bound states, for example, considering SW transformation to higher order, we expect the oscillations to damp.
The physics of such bound states also manifests itself in short-time enhancement of pair-boson correlations in $H_{\rm{nn}}$ and $H_{\rm{nnn}}$, plots Fig.~\ref{fig:HCBvsTrunc}~(a2)-(b2), although in these cases the hopping terms in the Hamiltonians do move the bound states, and we expect exponential decays on long time, similar to Fig.~\ref{fig:HCBvsTrunc}~(d2). 
Clearly, one has to consider short-distance physics details to understand these results in the pair-boson correlation function quantitatively.
Nevertheless, as far as observables discussed in the original spin model, since the pair-boson contributions are sub-dominant to the single-boson contributions, we expect the exponential decay behavior is robust under adding interactions to the simplified hard-core boson hopping Hamiltonian.

\section{A quick study of ``non-thermalizing" initial state \texorpdfstring{$|X\!+ \rangle$}{Lg}}
\label{sec:nonthermalizing}
In the same spin model, Ref.~\cite{Banuls2011} also found apparent ``non-thermalizing" behavior for the initial state $|X\!+ \rangle$.
This case shows much smaller oscillations than the weak thermalization case, but apparently observables approach non-thermal values at the longest simulation times.
While we do not have as clear picture of this case compared to the weak thermalization case, we will briefly discuss how far similar physical reasoning can take us in the non-thermalizing case.

First of all, the initial state $|X\!+ \rangle$ lands close to the top of the spectrum of the original Hamiltonian, Eq.~(\ref{H}).
Equivalently, it is close to the ground state of the Hamiltonian $\tilde{H} = -H$; this is the language we adopt since we are more used to think about ground states and low-energy excitations.
Here, we can develop a perturbative treatment starting with $J, h \ll |g|$, where the ground state is guaranteed to be close to $|X\!+ \rangle$.
Such a perturbative treatment is, in fact, fairly reasonable for the parameters of interest $J = 1, h = 0.5, g = -1.05$:
Indeed, $h$ is smaller than $g$, while for states with aligned spins the $+J \sigma_j^z \sigma_{j+1}^z$ terms in $\tilde{H}$ are frustrated.
We find that the $|X\!+ \rangle$ state has weight $|\langle \tilde{E}_0 |X\!+ \rangle|^2 \approx 23\%$ on the ground state of $\tilde{H}$ for $L=18$, which is smaller than in the weak thermalization case but is still a large weight.
The average energy density in the initial state is $\langle X\!+ | \tilde{H} | X\!+ \rangle/L - \tilde{E}_0/L \approx 0.28$ is somewhat larger than in the weak thermalization case.

In the perturbative picture, low-energy excitations are spins oriented in the $-\hat{x}$ direction (i.e., flipped compared to the ground state $|X\!+ \rangle$), with dispersion at leading order $\epsilon_k = 2|g| + 2J\cos(k)$.
The bottom of the quasiparticle band now lies at $k = \pi $ and can be quite close to the ground state, since $\epsilon_{k = \pi} = 2|g| - 2J$ becomes small for $J$ approaching $|g|$.
Our ED results indeed show a fairly small gap at $k = \pi$; the gap is likely smaller than $0.35$ in the thermodynamic limit and has strong even-odd effect on $L$ coinciding with whether the mesh of $k$-points contains $\pi$ or not.
This gap is an order of magnitude smaller than the quasiparticle gap of interest in the weak thermalization case.
Of relevance to the study of translationally invariant initial states, the gap to the lowest excitation with momentum $k = 0$ is larger.
This gap also has strong even-odd effect on $L$ and is likely smaller than $1.0$, which is almost four times smaller than in the weak thermalization case.
The lowest $k = 0$ excitation is likely a composite of two quasiparticles near the bottom of the band at momentum $\pi$, i.e., it is not simply a single spin-flip of the band $\epsilon_k = 2|g| + 2 J \cos(k)$ at $k = 0$.
Finally, the apparent velocity for the propagation of quantum correlations is two or more times larger than in the weak thermalization case, as judged from the observed much shorter recurrence times in our finite systems.

All of the above points to a more complex picture in terms of quasiparticles, which are likely moving faster but are also less sharp due to smaller gap and stronger mixing with multi-particle states.
Hence, our intuition is that the system should relax faster, which is indeed observed in Ref.~\cite{Banuls2011}.
However, we would also naively conclude that the system will approach the ``thermalized" state, contrary to what is observed in Ref.~\cite{Banuls2011} on the accessible simulation times.
It is possible that the thermalization does eventually happen, but the combined effects of the smallness of the gap and frustrating interactions produce a complex behavior on intermediate time scales.
At the same time, the quantum correlations spread more quickly, which limits the accessible simulation times before entanglement increases too much in the infinite-MPS study.
A very interesting possibility would be if this system does not thermalize in the conventional sense because of some emergent integral of motion, perhaps of the kind discussed in Refs.~\cite{Grover2014, Garrison2016} and in our discussion of the truncated SW picture earlier.
Alternatively, it could be that it thermalizes very slowly because of an approximate integral of motion.
Since we are unable to provide a more controlled understanding of the non-thermalizing behavior, we leave this as an interesting open problem.

\section{Conclusions}
\label{sec:concl}
We studied the origin and eventual fate of strong oscillations in specific quantum quenches in the non-integrable spin model, where the initial state has low energy density relative to the ground state.
By extrapolating our finite-size ED calculations, we were able to interpret the oscillation frequency as the quasiparticle creation energy.
We further used SW transformation to derive the effective Hamiltonian to have a better description of the quasiparticles at finite density.
The time evolution problem can be viewed as a quench from a dilute BEC state to an interacting hard-core boson Hamiltonian.
The oscillation signal mainly comes from the observables changing the particle number by one.

Inspired by the finite-size ED and perturbative SW calculations, we further simplified the problem by considering two specific BEC initial states and the hard-core boson Hamiltonian with hopping only.
This problem is interesting on its own even without the context of the spin problem we discussed.
We considered first the initial state as a simple product state analogous to the boson coherent state but under hard-core constraint. 
The other initial state was prepared as a topological superconductor in the Jordan-Wigner fermionic representation, which we argued has long range order in the bosonic representation and is qualitatively in the same BEC phase as the first state.
Furthermore, Wick's theorem is valid for this state, allowing us to obtain results for much larger systems and longer times via the Pfaffian method.

Incidentally, we discovered that under the condition Eq.~(\ref{muurho}), the topological superconductor wave function, when written in terms of bosons and projected into a sector with a fixed particle number, has amplitudes independent of the positions of the particles, thus becoming the initial BEC state studied in Refs.~\cite{Kormos2014}~and~\cite{Mazza2014}.
Since the restriction to fixed particle number is not important in the thermodynamic limit, in principle, one can study essentially the same quench problems but with the advantage of Wick's theorem.
The bosonic two-point correlation function hence reduces to an evaluation of a block Toeplitz determinant with $2 \times 2$ blocks.
Such block Toeplitz determinants are not as well studied, and we have not been able to obtain an analytical expression in our case (e.g., some of the calculation tricks in Ref.~\cite{Calabrese2012} were not directly applicable to our problem).
Finding such a compact expression for the correlation functions remains an interesting outstanding problem. 

Our numerical calculations strongly suggest that the BEC order parameter $\langle b_j^\dagger(t) \rangle$ decays exponentially with time.
Both the decay rate and the oscillation frequency depend on the boson density. 
We believe that the exponential decay originates from the non-local nature of the boson creation operator in terms of the JW fermions.
The non-locality of the operator excites quasiparticles at all momenta, whose interferences produce the exponential decay.
Using analogy with quenches from the ferromagnetic state in the quantum Ising chain \cite{Calabrese2012,Essler2016}, we conjectured the inverse lifetime as Eq.~(\ref{lifetime}), and hence obtained explicit density dependence of the lifetime. 
We also conjectured that the oscillation frequency depends on the density as Eq.~(\ref{omegarho}).
On the other hand, the pair-boson observable $\langle b_j^\dagger(t) b_{j+1}^\dagger(t) \rangle$ also has oscillations but with a $t^{-3/2}$ power-law decay, and the oscillation frequencies and power-law decay behavior are independent of the density in the initial state.

As an open problem for quenches from the BEC state, it will be interesting to study the validity of our conjectures for the decoherence time and oscillation frequency in the specific hard-core boson hopping model. 
Our calculation of the different decaying behaviors for single- and pair-boson observables can in principle be distinguished and verified in cold atom experiments such as the setting in Ref.~\cite{Greiner2002}, with non-zero hopping and much stronger on-site interaction. 
It will also be interesting to add integrability-preserving and integrability-breaking interactions to the hard-core boson hopping model and study how these affect the described behaviors.

Returning to the original spin system exhibiting weak thermalization, we made several approximations and simplifications when mapping this to the BEC quench problem.
In particular, we implicitly introduced an additional conserved quantity in our SW treatment by dropping $H_{\rm other}$ terms that connect different sectors.
Specifically, our truncated-SW Hamiltonian conserves the sector identity, or equivalently the total quasiparticle number.
An interesting possibility would be that such successive SW transformations developed to higher orders converge, implying true emergence of such integrals of motions in the translationally invariant system, in the spirit of Refs.~\cite{Grover2014,Garrison2016}. 
One can also view this SW transformation as a kind of renormalization, which pushes the effect of changing the excitation numbers to lower energy in the effective Hamiltonian, but at the expense of making the original observables more complex.
However, at present we do not know how to address this interesting possibility in thermodynamically large systems.

Fairly conservatively, we would expect that the dropped terms in the Hamiltonian would lead to improved thermalization, in the sense that the system is more generic and likely to thermalize at long times.
Since even without those dropped terms and with the simplified initial state structure, we showed that the oscillation signal still decays, we expect in the full problem the oscillations will decay in the long time limit.
Nevertheless, the decay becomes slower as we decrease the quasiparticle density and can be particularly slow at small density.
Therefore, we boldly conclude that the oscillation will decay eventually in the weak thermalization regime, and the apparently persistent oscillation is due to its slow decay rate as a result of the low density of the quasiparticles.

We would also like to mention a puzzling problem regarding the $|X-\rangle$ initial state studied in Ref.~\cite{Hastings2015}, which also shows the persistent oscillation behavior up to time $t=20$. 
This state is also on the lower end of the spectrum, but with three times higher energy density than the $|Z+\rangle$ state. 
Our naive estimation using the hard-core boson hopping model  would give us a much shorter decoherence time disagreeing with the infinite-MPS simulation. 
We suspect that to get a more quantitative agreement, we need to incorporate further details about the quasiparticle interactions, such as those discussed at the end of Sec.~\ref{subsec:qppic}. 
We will leave this for the future work.

\begin{acknowledgments}
We gratefully acknowledge M. Fisher, J. Garrison, S. Gopalakrishnan, R. Mishmash, G. Refael, and C. White for many helpful discussions and also thank S. Gopalakrishnan for bringing Ref.~\cite{McCoy1971} to our attention.
We are particularly grateful to M.-C. Ba\~{n}uls, I. Cirac, and M. Hastings for sharing their infinite-MPS data used in Fig.~\ref{fig:EDevolution} and for valuable comments on the paper and also for pointing us to Ref.~\cite{Hastings2015}.
We would also like to thank P. Calabrese and G. Delfino for bringing Refs.~\cite{Kormos2016, Delfino2014, Delfino2016} to our attention.
This work was supported by the NSF through Grants No. DMR-1206096 and No. DMR-1619696 and the Caltech Institute for Quantum Information and Matter, an NSF Physics Frontiers Center, with support of the Gordon and Betty Moore Foundation.

\end{acknowledgments}

\appendix

\section{Appendix A: Local Schrieffer-Wolf Transformation}
\label{app:SW}
We consider $H_0$, Eq.~(\ref{H0}), as our basic solvable Hamiltonian and treat $T$, Eq.~(\ref{T}), as our perturbation.
The latter can be decomposed as
\begin{equation}
T = T_{1,2} + T_{-1,-2} + T_{1,0} + T_{-1,0} + T_{1,-2} + T_{-1,2} ~,
\end{equation} 
with
\begin{eqnarray*}
T_{1,2}   &=& -g \sum_j P_{j-1}^\up \sigma_j^- P_{j+1}^\up ~, \\
T_{-1,-2} &=& -g \sum_j P_{j-1}^\up \sigma_j^+ P_{j+1}^\up ~, \\
T_{1,0}   &=& -g \sum_j \left( P_{j-1}^\up \sigma_j^- P_{j+1}^\dn + P_{j-1}^\dn \sigma_j^- P_{j+1}^\up \right) ~, \\
T_{-1,0}  &=& -g \sum_j \left( P_{j-1}^\up \sigma_j^+ P_{j+1}^\dn + P_{j-1}^\dn \sigma_j^+ P_{j+1}^\up \right) ~, \\
T_{1,-2}  &=& -g \sum_j P_{j-1}^\dn \sigma_j^- P_{j+1}^\dn ~, \\
T_{-1,2}  &=& -g \sum_j P_{j-1}^\dn \sigma_j^+ P_{j+1}^\dn ~. \\
\end{eqnarray*}
Each $T_{m,n}$ satisfies $[H_0, T_{m,n}] = 2 (mh + nJ) T_{m,n}$ and works like a generalized ladder operator on the energy levels of $H_0$.
Furthermore, $T_{-m,-n} = T_{m,n}^\dagger$.

We develop a perturbative local Schieffer-Wolff approach following Ref.~\cite{MacDonald1988}.
Consider a unitary transformation $e^{iS}$ with the generator 
\begin{equation}
iS = iS^{[1]} + iS^{[2]} + \dots ~,
\end{equation}
where $iS^{[k]}$ is of order $O(g^k)$.
We can expand the rotated Hamiltonian as 
\begin{eqnarray*}
H' &\equiv& e^{iS} H e^{-iS} = H_0 + T + [iS^{[1]}, H_0] \\
&+& [iS^{[1]}, T] + [iS^{2}, H_0] + \frac{1}{2}[iS^{[1]}, [iS^{[1]}, H_0]] + O(g^3) ~.
\end{eqnarray*}
The generators $iS^{[k]}$ are chosen order by order so as to eliminate the excitation-number-changing part of the previous order.
Specifically, we choose $iS^{[1]}$ such that
\begin{equation}
T + [iS^{[1]}, H_0] = 0 ~,
\end{equation}
with the solution
\begin{equation}
iS^{[1]} = \frac{T_{1,2} - T_{-1,-2}}{2h + 4J} + \frac{T_{1,0} - T_{-1,0}}{2h} + \frac{T_{1,-2} - T_{-1,2}}{2h - 4J} ~.
\end{equation}
To this order,
\begin{equation}
H' = H_0 + \frac{1}{2}[iS^{[1]}, T] + [iS^{[2]}, H_0] + O(g^3) ~,
\end{equation}
where the second term contains both excitation-number-preserving terms (i.e., sector-diagonal terms) and excitation-number-changing terms (i.e., sector-off-diagonal terms), now in second order.
To eliminate the excitation-number-changing-terms to this order, we choose $iS^{[2]}$ such that 
\begin{equation}
\left( \frac{1}{2}[iS^{[1]}, T] \right)_{\rm sector-off-diag} + [iS^{[2]}, H_0] = 0 ~,
\label{iS2eq}
\end{equation}
giving us
\begin{equation}
H' = H_0 + \left( \frac{1}{2}[iS^{[1]}, T] \right)_{\rm sector-diag} + O(g^3) ~.
\end{equation}
This is the Hamiltonian quoted in the main text, Eq.~(\ref{Hprime}), with $H_{\rm other}$ containing terms of order $O(g^3)$.

While we do not need explicit $iS^{[2]}$ to determine the effective Hamiltonian to this order, we do use it when rotating the operators and the initial state in Sec.~\ref{truncatedSW} in calculations leading to Fig.~\ref{fig:InitialState}.
Hence we quote the solution to Eq.~(\ref{iS2eq}):
\begin{eqnarray}
iS^{[2]} &=&
\left( \frac{1}{4h} - \frac{1}{8J-4h} \right) \frac{[T_{1,0}, T_{-1,2}] - \Hc}{4J} \nonumber\\
&+& \left( \frac{1}{8J-4h} + \frac{1}{4h} \right) \frac{[T_{1,-2}, T_{1,0}] - \Hc}{4J-4h} \nonumber\\
&+& \left( \frac{1}{4h} - \frac{1}{8J+4h} \right) \frac{[T_{1,0}, T_{1,2}] - \Hc}{4J+4h} \nonumber\\
&+& \left( \frac{1}{4h} + \frac{1}{8J+4h} \right) \frac{[T_{-1,-2}, T_{1,0}] - \Hc}{4J} \nonumber\\
&+& \left( \frac{1}{8J-4h} - \frac{1}{8J+4h} \right) \frac{[T_{-1,2}, T_{1,2}] - \Hc}{8J} \nonumber\\
&+& \left( \frac{1}{8J+4h} + \frac{1}{8J-4h} \right) \frac{[T_{-1,-2}, T_{-1,2}] - \Hc}{4h} ~.
\end{eqnarray}
While in principle we can continue to obtain higher order effective Hamiltonians, the form of the previous order will not be changed.
Note that in the main text, for the purpose of demonstrating the SW picture, we used the perturbatively truncated $H'$, while we rotated the operators and the initial state using $\exp(iS^{[1]} + iS^{[2]})$ (i.e., without further expansion of the exponential).  
This was actually easier to implement and also guarantees that the operators and initial state are rotated by a unitary and will agree with perturbative treatment to this order (when studying dynamics, this also assumes appropriately small elapsed time).
We did not use exact unitary rotation of the Hamiltonian since then the rotated problem would be unitarily equivalent to the original problem, with no new information.
We expect that using truncated Hamiltonian, which in addition conserves excitation number, should make it only more difficult to thermalize, and we indeed do not observe any faster thermalization.

For completeness, we also present below the effective observables to first order, even though we did not explicitly use them in the main text.
To obtain the observables in the rotated picture, we use the formula $\hat{O}' = e^{iS} \hat{O} e^{-iS} = \hat{O} + [iS^{[1]}, \hat{O}] + O(g^2)$. 
With $\hat{O} = \sigma_j^x$, $\sigma_j^y$ and, $\sigma_j^z$, we have
\begin{widetext}
\begin{eqnarray}
\label{sigmaxprime}
(\sigma_j^x)' &\approx&  \sigma_j^x + \frac{g}{2h}(\sigma_{j-2}^z \sigma_{j-1}^y \sigma_j^y + \sigma_j^{y} \sigma_{j+1}^y\sigma_{j+2}^z + 2 P^{\uparrow}_{j-1} \sigma_j^z P^{\downarrow}_{j+1} + 2P^{\downarrow}_{j-1} \sigma_j^z P^{\uparrow}_{j+1}) \\
&-& \frac{g}{4J-2h}(P^{\downarrow}_{j-2}\sigma^y_{j-1}\sigma^{y}_{j}+\sigma^{y}_{j}\sigma^{y}_{j+1}P^{\downarrow}_{j+2}+2P^{\downarrow}_{j-1}\sigma_j^{z}p^{\downarrow}_{j+1}) 
+\frac{g}{4J+2h}(P^{\uparrow}_{j-2}\sigma^{y}_{j-1}\sigma^{y}_{j}+\sigma^{y}_{j}\sigma^{y}_{j+1}P^{\uparrow}_{j+2}-2P^{\uparrow}_{j-1}\sigma^{z}_jP^{\uparrow}_{j+1}) \nonumber \\
\label{sigmayprime}
(\sigma_j^y)' &\approx& \sigma_j^y - \frac{g}{2h}(\sigma_{j-2}^z \sigma_{j-1}^y \sigma_j^x + \sigma_j^x \sigma_{j+1}^y \sigma_{j+2}^z)
+ \frac{g}{4J-2h}(P^{\downarrow}_{j-2} \sigma^y_{j-1} \sigma^x_j + \sigma^x_j \sigma^y_{j+1} P^{\downarrow}_{j+2}) \\
&-& \frac{g}{4J+2h}(P^{\uparrow}_{j-2}\sigma^{y}_{j-1}\sigma^x_j + \sigma^y_{j+1} P^{\uparrow}_{j+2}) ~, \nonumber
\end{eqnarray}
\end{widetext}
and $(\sigma_j^z)'$ as in Eq.~(\ref{sigmazprime}).

\section{Calculation of \texorpdfstring{$\langle b_j^\dagger(t) b_{j+1}^\dagger(t) \rangle$}{Lg}, Eq.~(\ref{bbeq}).}
\label{app:BB}
In this Appendix, we present details of the calculation of $\Upsilon(t) \equiv \langle b_j^\dagger(t) b_{j+1}^\dagger(t) \rangle$ for the initial product BEC state.
We first obtain a closed-form expression for the observable in a finite system of length $L$ and then derive the thermodynamic limit Eq.~(\ref{bbeq}).
Since the Jordan-Wigner fermions have different boundary conditions in the even- and odd-number-parity sectors, when calculating the time evolution we split
\begin{eqnarray}
\Upsilon(t) &=& \Upsilon_{\rm even}(t) + \Upsilon_{\rm odd}(t) ~, \\
\Upsilon_{\rm even (odd)}(t) &\equiv & \langle P_{\rm even (odd)} b_j^\dagger(t) b_{j+1}^\dagger(t) P_{\rm even (odd)} \rangle ~,\nonumber \\
\end{eqnarray}
where $P_{\rm even (odd)}$ is the projector to the even- (odd-)number-parity sector.

We can use translational invariance to consider instead $\langle \sum_j b_j^\dagger(t) b_{j+1}^\dagger(t) \rangle / L$.
Keeping in mind implicit surrounding sector projectors, we can express
\begin{eqnarray}
&& \frac{1}{L} \sum_j b_j^\dagger(t) b_{j+1}^\dagger(t) =
\frac{1}{L} \sum_k c_k^\dagger c_{-k}^\dagger e^{i k} e^{i 2\epsilon_k t} \\
&& = \frac{1}{L^2} \sum_k e^{i k} e^{i 2\epsilon_k t} \sum_{j, j'} c_j^\dagger c_{j'}^\dagger e^{i k (j-j')} \\
&& = \frac{2}{L^2} \sum_k \sin(k) e^{i 2\epsilon_k t} \sum_{j < j'} \sin[k (j'-j)] c_j^\dagger c_{j'}^\dagger ~,
\label{bbsum}
\end{eqnarray}
where momenta $k$ should be taken appropriately for each number-parity sector.
We also used $\epsilon_{-k} = \epsilon_k$ and kept only the surviving total even part in $k$ in the last line.

Restoring the sector projectors, we now need to evaluate $\langle P_{\rm even (odd)} c_j^\dagger c_{j'}^\dagger P_{\rm even (odd)} \rangle$ in the initial product BEC state Eq.~(\ref{hcbBEC}).
We can express the projectors as $P_{\rm even (odd)} = (1 \pm e^{i \pi N_{\rm tot}})/2$.
Writing the fermionic operators in terms of the bosonic operators gives $c_j^\dagger c_{j'}^\dagger = b_j^\dagger \left(\prod_{s=j+1}^{j'-1} e^{i\pi n_s} \right) b_{j'}^\dagger$, where we assumed $j < j'$.
We can easily evaluate expectation values in the product BEC state and obtain
\begin{equation*}
\langle P_{\rm even (odd)} c_j^\dagger c_{j'}^\dagger P_{\rm even (odd)} \rangle \!=\! (\beta^* \alpha)^2
\frac{\eta^{j'-j-1} \pm \eta^{L-(j'-j+1)}}{2},
\end{equation*}
where we defined $\eta \equiv \langle e^{i\pi n_s} \rangle = |\alpha|^2 - |\beta|^2$.

When calculating the expectation value of Eq.~(\ref{bbsum}), the summation over $j < j'$ actually contains only a function of $j'-j$.
For each separation $\ell = j'-j$, there are $L - \ell$ identical terms to be summed.
Therefore, we now need to calculate
\begin{eqnarray}
&& \sum_{\ell=1}^{L-1} (L-\ell) \sin(k\ell) (\eta^{\ell-1} \pm \eta^{L-\ell-1}) \nonumber \\
&& = \pm \sum_{s=1}^{L-1} s \sin(ks) (\eta^{L-s-1} \pm \eta^{s-1}) \nonumber \\
&& = \sum_{s=1}^{L-1} {\rm Im} \left[ s e^{i k s} (\pm \eta^{L-s-1} + \eta^{s-1}) \right] ~,
\end{eqnarray}
where in the second line we changed the summation variable to $s \equiv L - \ell$ and also used that $k L = 2\pi m + \pi$ or $k L = 2\pi m$, $m \in \mathbb{Z}$, in the even- or odd-parity sectors respectively (upper and lower signs respectively).

The summation can be done exactly using the formula
\begin{equation}
\sum_{s=1}^{L-1} s a^s = a \frac{\partial}{\partial a} \sum_{s=1}^{L-1} a^s = \frac{a (1 - a^L)}{(1-a)^2} - \frac{L a^L}{1-a} ~,
\end{equation}
applying it with $a = e^{ik} \eta^{-1}$ and $a = e^{ik} \eta$ for the first and second parts respectively.
Putting everything together, we find
\begin{eqnarray*}
&& \Upsilon_{\rm even (odd)}(t) = (\beta^* \alpha)^2
\frac{1}{L^2} \sum_k \sin(k) e^{i 2\epsilon_k t} \\
&& {\rm Im}\left[ \frac{e^{ik}(1 \pm \eta^L)}{(\eta - e^{ik})^2} + \frac{L}{\eta - e^{ik}} 
+ \frac{e^{ik} (1 \pm  \eta^L)}{(1 - e^{ik} \eta)^2} \pm \frac{L \eta^{L-1}}{1 - e^{ik} \eta} \right] ~,
\end{eqnarray*}
where we again used $e^{i k L} = \mp$ in the even (odd) sectors respectively.
The finite-site result can be easily summed numerically at this point for any $L$, which is how we obtained the corresponding data in the main text.
We can also easily take the thermodynamic limit, remembering that $|\eta| < 1$.
In particular, we see that in the thermodynamic limit, the contributions from the even- and odd-number-parity sectors are the same, and we obtain Eq.~(\ref{bbeq}) in the main text.


\begin{thebibliography}{53}%
\makeatletter
\providecommand \@ifxundefined [1]{%
 \@ifx{#1\undefined}
}%
\providecommand \@ifnum [1]{%
 \ifnum #1\expandafter \@firstoftwo
 \else \expandafter \@secondoftwo
 \fi
}%
\providecommand \@ifx [1]{%
 \ifx #1\expandafter \@firstoftwo
 \else \expandafter \@secondoftwo
 \fi
}%
\providecommand \natexlab [1]{#1}%
\providecommand \enquote  [1]{``#1''}%
\providecommand \bibnamefont  [1]{#1}%
\providecommand \bibfnamefont [1]{#1}%
\providecommand \citenamefont [1]{#1}%
\providecommand \href@noop [0]{\@secondoftwo}%
\providecommand \href [0]{\begingroup \@sanitize@url \@href}%
\providecommand \@href[1]{\@@startlink{#1}\@@href}%
\providecommand \@@href[1]{\endgroup#1\@@endlink}%
\providecommand \@sanitize@url [0]{\catcode `\\12\catcode `\$12\catcode
  `\&12\catcode `\#12\catcode `\^12\catcode `\_12\catcode `\%12\relax}%
\providecommand \@@startlink[1]{}%
\providecommand \@@endlink[0]{}%
\providecommand \url  [0]{\begingroup\@sanitize@url \@url }%
\providecommand \@url [1]{\endgroup\@href {#1}{\urlprefix }}%
\providecommand \urlprefix  [0]{URL }%
\providecommand \Eprint [0]{\href }%
\providecommand \doibase [0]{http://dx.doi.org/}%
\providecommand \selectlanguage [0]{\@gobble}%
\providecommand \bibinfo  [0]{\@secondoftwo}%
\providecommand \bibfield  [0]{\@secondoftwo}%
\providecommand \translation [1]{[#1]}%
\providecommand \BibitemOpen [0]{}%
\providecommand \bibitemStop [0]{}%
\providecommand \bibitemNoStop [0]{.\EOS\space}%
\providecommand \EOS [0]{\spacefactor3000\relax}%
\providecommand \BibitemShut  [1]{\csname bibitem#1\endcsname}%
\let\auto@bib@innerbib\@empty
\bibitem [{\citenamefont {Srednicki}(1994)}]{Srednicki1994}%
  \BibitemOpen
  \bibfield  {author} {\bibinfo {author} {\bibfnamefont {M.}~\bibnamefont
  {Srednicki}},\ }\href {\doibase 10.1103/PhysRevE.50.888} {\bibfield
  {journal} {\bibinfo  {journal} {Phys. Rev. E}\ }\textbf {\bibinfo {volume}
  {50}},\ \bibinfo {pages} {888} (\bibinfo {year} {1994})}\BibitemShut
  {NoStop}%
\bibitem [{\citenamefont {Deutsch}(1991)}]{Deutsch1991}%
  \BibitemOpen
  \bibfield  {author} {\bibinfo {author} {\bibfnamefont {J.~M.}\ \bibnamefont
  {Deutsch}},\ }\href {\doibase 10.1103/PhysRevA.43.2046} {\bibfield  {journal}
  {\bibinfo  {journal} {Phys. Rev. A}\ }\textbf {\bibinfo {volume} {43}},\
  \bibinfo {pages} {2046} (\bibinfo {year} {1991})}\BibitemShut {NoStop}%
\bibitem [{\citenamefont {Kim}\ \emph {et~al.}(2014)\citenamefont {Kim},
  \citenamefont {Ikeda},\ and\ \citenamefont {Huse}}]{Kim2014}%
  \BibitemOpen
  \bibfield  {author} {\bibinfo {author} {\bibfnamefont {H.}~\bibnamefont
  {Kim}}, \bibinfo {author} {\bibfnamefont {T.~N.}\ \bibnamefont {Ikeda}}, \
  and\ \bibinfo {author} {\bibfnamefont {D.~A.}\ \bibnamefont {Huse}},\ }\href
  {\doibase 10.1103/PhysRevE.90.052105} {\bibfield  {journal} {\bibinfo
  {journal} {Phys. Rev. E}\ }\textbf {\bibinfo {volume} {90}},\ \bibinfo
  {pages} {052105} (\bibinfo {year} {2014})}\BibitemShut {NoStop}%
\bibitem [{\citenamefont {Beugeling}\ \emph {et~al.}(2014)\citenamefont
  {Beugeling}, \citenamefont {Moessner},\ and\ \citenamefont
  {Haque}}]{beugeling_finite-size_2014}%
  \BibitemOpen
  \bibfield  {author} {\bibinfo {author} {\bibfnamefont {W.}~\bibnamefont
  {Beugeling}}, \bibinfo {author} {\bibfnamefont {R.}~\bibnamefont {Moessner}},
  \ and\ \bibinfo {author} {\bibfnamefont {M.}~\bibnamefont {Haque}},\ }\href
  {\doibase 10.1103/PhysRevE.89.042112} {\bibfield  {journal} {\bibinfo
  {journal} {Phys. Rev. E}\ }\textbf {\bibinfo {volume} {89}},\ \bibinfo
  {pages} {042112} (\bibinfo {year} {2014})}\BibitemShut {NoStop}%
\bibitem [{\citenamefont {Garrison}\ and\ \citenamefont
  {Grover}()}]{Garrison2015}%
  \BibitemOpen
  \bibfield  {author} {\bibinfo {author} {\bibfnamefont {J.~R.}\ \bibnamefont
  {Garrison}}\ and\ \bibinfo {author} {\bibfnamefont {T.}~\bibnamefont
  {Grover}},\ }\href {http://arxiv.org/abs/1503.00729} {\ }\Eprint
  {http://arxiv.org/abs/1503.00729} {arXiv:1503.00729} \BibitemShut {NoStop}%
\bibitem [{\citenamefont {Beugeling}\ \emph
  {et~al.}(2015{\natexlab{a}})\citenamefont {Beugeling}, \citenamefont
  {Andreanov},\ and\ \citenamefont {Haque}}]{beugeling_global_2015}%
  \BibitemOpen
  \bibfield  {author} {\bibinfo {author} {\bibfnamefont {W.}~\bibnamefont
  {Beugeling}}, \bibinfo {author} {\bibfnamefont {A.}~\bibnamefont
  {Andreanov}}, \ and\ \bibinfo {author} {\bibfnamefont {M.}~\bibnamefont
  {Haque}},\ }\href {\doibase 10.1088/1742-5468/2015/02/P02002} {\bibfield
  {journal} {\bibinfo  {journal} {J. Stat. Mech.}\ }\textbf {\bibinfo {volume}
  {2015}},\ \bibinfo {pages} {P02002} (\bibinfo {year}
  {2015}{\natexlab{a}})}\BibitemShut {NoStop}%
\bibitem [{\citenamefont {Mondaini}\ \emph {et~al.}(2016)\citenamefont
  {Mondaini}, \citenamefont {Fratus}, \citenamefont {Srednicki},\ and\
  \citenamefont {Rigol}}]{Mondaini2016}%
  \BibitemOpen
  \bibfield  {author} {\bibinfo {author} {\bibfnamefont {R.}~\bibnamefont
  {Mondaini}}, \bibinfo {author} {\bibfnamefont {K.~R.}\ \bibnamefont
  {Fratus}}, \bibinfo {author} {\bibfnamefont {M.}~\bibnamefont {Srednicki}}, \
  and\ \bibinfo {author} {\bibfnamefont {M.}~\bibnamefont {Rigol}},\ }\href
  {\doibase 10.1103/PhysRevE.93.032104} {\bibfield  {journal} {\bibinfo
  {journal} {Phys. Rev. E}\ }\textbf {\bibinfo {volume} {93}},\ \bibinfo
  {pages} {032104} (\bibinfo {year} {2016})}\BibitemShut {NoStop}%
\bibitem [{\citenamefont {Rigol}\ \emph {et~al.}(2008)\citenamefont {Rigol},
  \citenamefont {Dunjko},\ and\ \citenamefont {Olshanii}}]{Rigol2008}%
  \BibitemOpen
  \bibfield  {author} {\bibinfo {author} {\bibfnamefont {M.}~\bibnamefont
  {Rigol}}, \bibinfo {author} {\bibfnamefont {V.}~\bibnamefont {Dunjko}}, \
  and\ \bibinfo {author} {\bibfnamefont {M.}~\bibnamefont {Olshanii}},\ }\href
  {http://dx.doi.org/10.1038/nature06838} {\bibfield  {journal} {\bibinfo
  {journal} {Nature}\ }\textbf {\bibinfo {volume} {452}},\ \bibinfo {pages}
  {854} (\bibinfo {year} {2008})}\BibitemShut {NoStop}%
\bibitem [{\citenamefont {Greiner}\ \emph {et~al.}(2002)\citenamefont
  {Greiner}, \citenamefont {Mandel}, \citenamefont {H{\"{a}}nsch},\ and\
  \citenamefont {Bloch}}]{Greiner2002}%
  \BibitemOpen
  \bibfield  {author} {\bibinfo {author} {\bibfnamefont {M.}~\bibnamefont
  {Greiner}}, \bibinfo {author} {\bibfnamefont {O.}~\bibnamefont {Mandel}},
  \bibinfo {author} {\bibfnamefont {T.~W.}\ \bibnamefont {H{\"{a}}nsch}}, \
  and\ \bibinfo {author} {\bibfnamefont {I.}~\bibnamefont {Bloch}},\ }\href
  {\doibase 10.1038/nature00968} {\bibfield  {journal} {\bibinfo  {journal}
  {Nature}\ }\textbf {\bibinfo {volume} {419}},\ \bibinfo {pages} {51}
  (\bibinfo {year} {2002})}\BibitemShut {NoStop}%
\bibitem [{\citenamefont {Kinoshita}\ \emph {et~al.}(2006)\citenamefont
  {Kinoshita}, \citenamefont {Wenger},\ and\ \citenamefont
  {Weiss}}]{Kinoshita2006}%
  \BibitemOpen
  \bibfield  {author} {\bibinfo {author} {\bibfnamefont {T.}~\bibnamefont
  {Kinoshita}}, \bibinfo {author} {\bibfnamefont {T.}~\bibnamefont {Wenger}}, \
  and\ \bibinfo {author} {\bibfnamefont {D.~S.}\ \bibnamefont {Weiss}},\ }\href
  {\doibase 10.1038/nature04693} {\bibfield  {journal} {\bibinfo  {journal}
  {Nature}\ }\textbf {\bibinfo {volume} {440}},\ \bibinfo {pages} {900}
  (\bibinfo {year} {2006})}\BibitemShut {NoStop}%
\bibitem [{\citenamefont {Hofferberth}\ \emph {et~al.}(2007)\citenamefont
  {Hofferberth}, \citenamefont {Lesanovsky}, \citenamefont {Fischer},
  \citenamefont {Schumm},\ and\ \citenamefont
  {Schmiedmayer}}]{Hofferberth2007}%
  \BibitemOpen
  \bibfield  {author} {\bibinfo {author} {\bibfnamefont {S.}~\bibnamefont
  {Hofferberth}}, \bibinfo {author} {\bibfnamefont {I.}~\bibnamefont
  {Lesanovsky}}, \bibinfo {author} {\bibfnamefont {B.}~\bibnamefont {Fischer}},
  \bibinfo {author} {\bibfnamefont {T.}~\bibnamefont {Schumm}}, \ and\ \bibinfo
  {author} {\bibfnamefont {J.}~\bibnamefont {Schmiedmayer}},\ }\href
  {http://www.nature.com/doifinder/10.1038/nature06149} {\bibfield  {journal}
  {\bibinfo  {journal} {Nature}\ }\textbf {\bibinfo {volume} {449}},\ \bibinfo
  {pages} {324} (\bibinfo {year} {2007})}\BibitemShut {NoStop}%
\bibitem [{\citenamefont {Trotzky}\ \emph {et~al.}(2012)\citenamefont
  {Trotzky}, \citenamefont {Chen}, \citenamefont {Flesch}, \citenamefont
  {McCulloch}, \citenamefont {Schollwock}, \citenamefont {Eisert},\ and\
  \citenamefont {Bloch}}]{Trotzky2012}%
  \BibitemOpen
  \bibfield  {author} {\bibinfo {author} {\bibfnamefont {S.}~\bibnamefont
  {Trotzky}}, \bibinfo {author} {\bibfnamefont {Y.-A.}\ \bibnamefont {Chen}},
  \bibinfo {author} {\bibfnamefont {A.}~\bibnamefont {Flesch}}, \bibinfo
  {author} {\bibfnamefont {I.~P.}\ \bibnamefont {McCulloch}}, \bibinfo {author}
  {\bibfnamefont {U.}~\bibnamefont {Schollwock}}, \bibinfo {author}
  {\bibfnamefont {J.}~\bibnamefont {Eisert}}, \ and\ \bibinfo {author}
  {\bibfnamefont {I.}~\bibnamefont {Bloch}},\ }\href
  {http://dx.doi.org/10.1038/nphys2232} {\bibfield  {journal} {\bibinfo
  {journal} {Nat. Phys.}\ }\textbf {\bibinfo {volume} {8}},\ \bibinfo {pages}
  {325} (\bibinfo {year} {2012})}\BibitemShut {NoStop}%
\bibitem [{\citenamefont {Cheneau}\ \emph {et~al.}(2012)\citenamefont
  {Cheneau}, \citenamefont {Barmettler}, \citenamefont {Poletti}, \citenamefont
  {Endres}, \citenamefont {Schau{\ss}}, \citenamefont {Fukuhara}, \citenamefont
  {Gross}, \citenamefont {Bloch}, \citenamefont {Kollath},\ and\ \citenamefont
  {Kuhr}}]{Cheneau2012}%
  \BibitemOpen
  \bibfield  {author} {\bibinfo {author} {\bibfnamefont {M.}~\bibnamefont
  {Cheneau}}, \bibinfo {author} {\bibfnamefont {P.}~\bibnamefont {Barmettler}},
  \bibinfo {author} {\bibfnamefont {D.}~\bibnamefont {Poletti}}, \bibinfo
  {author} {\bibfnamefont {M.}~\bibnamefont {Endres}}, \bibinfo {author}
  {\bibfnamefont {P.}~\bibnamefont {Schau{\ss}}}, \bibinfo {author}
  {\bibfnamefont {T.}~\bibnamefont {Fukuhara}}, \bibinfo {author}
  {\bibfnamefont {C.}~\bibnamefont {Gross}}, \bibinfo {author} {\bibfnamefont
  {I.}~\bibnamefont {Bloch}}, \bibinfo {author} {\bibfnamefont
  {C.}~\bibnamefont {Kollath}}, \ and\ \bibinfo {author} {\bibfnamefont
  {S.}~\bibnamefont {Kuhr}},\ }\href {\doibase 10.1038/nature10748} {\bibfield
  {journal} {\bibinfo  {journal} {Nature}\ }\textbf {\bibinfo {volume} {481}},\
  \bibinfo {pages} {484} (\bibinfo {year} {2012})}\BibitemShut {NoStop}%
\bibitem [{\citenamefont {Gring}\ \emph {et~al.}(2012)\citenamefont {Gring},
  \citenamefont {Kuhnert}, \citenamefont {Langen}, \citenamefont {Kitagawa},
  \citenamefont {Rauer}, \citenamefont {Schreitl}, \citenamefont {Mazets},
  \citenamefont {Smith}, \citenamefont {Demler},\ and\ \citenamefont
  {Schmiedmayer}}]{Gring2012}%
  \BibitemOpen
  \bibfield  {author} {\bibinfo {author} {\bibfnamefont {M.}~\bibnamefont
  {Gring}}, \bibinfo {author} {\bibfnamefont {M.}~\bibnamefont {Kuhnert}},
  \bibinfo {author} {\bibfnamefont {T.}~\bibnamefont {Langen}}, \bibinfo
  {author} {\bibfnamefont {T.}~\bibnamefont {Kitagawa}}, \bibinfo {author}
  {\bibfnamefont {B.}~\bibnamefont {Rauer}}, \bibinfo {author} {\bibfnamefont
  {M.}~\bibnamefont {Schreitl}}, \bibinfo {author} {\bibfnamefont
  {I.}~\bibnamefont {Mazets}}, \bibinfo {author} {\bibfnamefont {D.~A.}\
  \bibnamefont {Smith}}, \bibinfo {author} {\bibfnamefont {E.}~\bibnamefont
  {Demler}}, \ and\ \bibinfo {author} {\bibfnamefont {J.}~\bibnamefont
  {Schmiedmayer}},\ }\href {\doibase 10.1126/science.1224953} {\bibfield
  {journal} {\bibinfo  {journal} {Science}\ }\textbf {\bibinfo {volume}
  {337}},\ \bibinfo {pages} {1318} (\bibinfo {year} {2012})}\BibitemShut
  {NoStop}%
\bibitem [{\citenamefont {Rigol}\ \emph {et~al.}(2007)\citenamefont {Rigol},
  \citenamefont {Dunjko}, \citenamefont {Yurovsky},\ and\ \citenamefont
  {Olshanii}}]{Rigol2007}%
  \BibitemOpen
  \bibfield  {author} {\bibinfo {author} {\bibfnamefont {M.}~\bibnamefont
  {Rigol}}, \bibinfo {author} {\bibfnamefont {V.}~\bibnamefont {Dunjko}},
  \bibinfo {author} {\bibfnamefont {V.}~\bibnamefont {Yurovsky}}, \ and\
  \bibinfo {author} {\bibfnamefont {M.}~\bibnamefont {Olshanii}},\ }\href
  {\doibase 10.1103/PhysRevLett.98.050405} {\bibfield  {journal} {\bibinfo
  {journal} {Phys. Rev. Lett.}\ }\textbf {\bibinfo {volume} {98}},\ \bibinfo
  {pages} {050405} (\bibinfo {year} {2007})}\BibitemShut {NoStop}%
\bibitem [{\citenamefont {Flesch}\ \emph {et~al.}(2008)\citenamefont {Flesch},
  \citenamefont {Cramer}, \citenamefont {McCulloch}, \citenamefont
  {Schollw{\"{o}}ck},\ and\ \citenamefont {Eisert}}]{Flesch2008a}%
  \BibitemOpen
  \bibfield  {author} {\bibinfo {author} {\bibfnamefont {A.}~\bibnamefont
  {Flesch}}, \bibinfo {author} {\bibfnamefont {M.}~\bibnamefont {Cramer}},
  \bibinfo {author} {\bibfnamefont {I.~P.}\ \bibnamefont {McCulloch}}, \bibinfo
  {author} {\bibfnamefont {U.}~\bibnamefont {Schollw{\"{o}}ck}}, \ and\
  \bibinfo {author} {\bibfnamefont {J.}~\bibnamefont {Eisert}},\ }\href
  {\doibase 10.1103/PhysRevA.78.033608} {\bibfield  {journal} {\bibinfo
  {journal} {Phys. Rev. A}\ }\textbf {\bibinfo {volume} {78}},\ \bibinfo
  {pages} {033608} (\bibinfo {year} {2008})}\BibitemShut {NoStop}%
\bibitem [{\citenamefont {Beugeling}\ \emph
  {et~al.}(2015{\natexlab{b}})\citenamefont {Beugeling}, \citenamefont
  {Moessner},\ and\ \citenamefont {Haque}}]{beugeling_off-diagonal_2015}%
  \BibitemOpen
  \bibfield  {author} {\bibinfo {author} {\bibfnamefont {W.}~\bibnamefont
  {Beugeling}}, \bibinfo {author} {\bibfnamefont {R.}~\bibnamefont {Moessner}},
  \ and\ \bibinfo {author} {\bibfnamefont {M.}~\bibnamefont {Haque}},\ }\href
  {\doibase 10.1103/PhysRevE.91.012144} {\bibfield  {journal} {\bibinfo
  {journal} {Phys. Rev. E}\ }\textbf {\bibinfo {volume} {91}},\ \bibinfo
  {pages} {012144} (\bibinfo {year} {2015}{\natexlab{b}})}\BibitemShut
  {NoStop}%
\bibitem [{\citenamefont {Collura}\ \emph {et~al.}(2015)\citenamefont
  {Collura}, \citenamefont {Calabrese},\ and\ \citenamefont
  {Essler}}]{Collura2015}%
  \BibitemOpen
  \bibfield  {author} {\bibinfo {author} {\bibfnamefont {M.}~\bibnamefont
  {Collura}}, \bibinfo {author} {\bibfnamefont {P.}~\bibnamefont {Calabrese}},
  \ and\ \bibinfo {author} {\bibfnamefont {F.~H.~L.}\ \bibnamefont {Essler}},\
  }\href {\doibase 10.1103/PhysRevB.92.125131} {\bibfield  {journal} {\bibinfo
  {journal} {Phys. Rev. B}\ }\textbf {\bibinfo {volume} {92}},\ \bibinfo
  {pages} {125131} (\bibinfo {year} {2015})}\BibitemShut {NoStop}%
\bibitem [{\citenamefont {Mazza}\ \emph {et~al.}(2016)\citenamefont {Mazza},
  \citenamefont {St\'ephan}, \citenamefont {Canovi}, \citenamefont {Alba},
  \citenamefont {Brockmann},\ and\ \citenamefont {Haque}}]{mazza_overlap_2016}%
  \BibitemOpen
  \bibfield  {author} {\bibinfo {author} {\bibfnamefont {P.~P.}\ \bibnamefont
  {Mazza}}, \bibinfo {author} {\bibfnamefont {J.-M.}\ \bibnamefont
  {St\'ephan}}, \bibinfo {author} {\bibfnamefont {E.}~\bibnamefont {Canovi}},
  \bibinfo {author} {\bibfnamefont {V.}~\bibnamefont {Alba}}, \bibinfo {author}
  {\bibfnamefont {M.}~\bibnamefont {Brockmann}}, \ and\ \bibinfo {author}
  {\bibfnamefont {M.}~\bibnamefont {Haque}},\ }\href
  {http://stacks.iop.org/1742-5468/2016/i=1/a=013104} {\bibfield  {journal}
  {\bibinfo  {journal} {J. Stat. Mech.}\ }\textbf {\bibinfo {volume} {2016}},\
  \bibinfo {pages} {013104} (\bibinfo {year} {2016})}\BibitemShut {NoStop}%
\bibitem [{\citenamefont {Rossini}\ \emph {et~al.}(2010)\citenamefont
  {Rossini}, \citenamefont {Suzuki}, \citenamefont {Mussardo}, \citenamefont
  {Santoro},\ and\ \citenamefont {Silva}}]{Rossini2010}%
  \BibitemOpen
  \bibfield  {author} {\bibinfo {author} {\bibfnamefont {D.}~\bibnamefont
  {Rossini}}, \bibinfo {author} {\bibfnamefont {S.}~\bibnamefont {Suzuki}},
  \bibinfo {author} {\bibfnamefont {G.}~\bibnamefont {Mussardo}}, \bibinfo
  {author} {\bibfnamefont {G.~E.}\ \bibnamefont {Santoro}}, \ and\ \bibinfo
  {author} {\bibfnamefont {A.}~\bibnamefont {Silva}},\ }\href
  {http://link.aps.org/doi/10.1103/PhysRevB.82.144302} {\bibfield  {journal}
  {\bibinfo  {journal} {Phys. Rev. B}\ }\textbf {\bibinfo {volume} {82}},\
  \bibinfo {pages} {144302} (\bibinfo {year} {2010})}\BibitemShut {NoStop}%
\bibitem [{\citenamefont {Calabrese}\ \emph {et~al.}(2011)\citenamefont
  {Calabrese}, \citenamefont {Essler},\ and\ \citenamefont
  {Fagotti}}]{Calabrese2011}%
  \BibitemOpen
  \bibfield  {author} {\bibinfo {author} {\bibfnamefont {P.}~\bibnamefont
  {Calabrese}}, \bibinfo {author} {\bibfnamefont {F.~H.~L.}\ \bibnamefont
  {Essler}}, \ and\ \bibinfo {author} {\bibfnamefont {M.}~\bibnamefont
  {Fagotti}},\ }\href {\doibase 10.1103/PhysRevLett.106.227203} {\bibfield
  {journal} {\bibinfo  {journal} {Phys. Rev. Lett.}\ }\textbf {\bibinfo
  {volume} {106}},\ \bibinfo {pages} {227203} (\bibinfo {year}
  {2011})}\BibitemShut {NoStop}%
\bibitem [{\citenamefont {Calabrese}\ \emph
  {et~al.}(2012{\natexlab{a}})\citenamefont {Calabrese}, \citenamefont
  {Essler},\ and\ \citenamefont {Fagotti}}]{Calabrese2012}%
  \BibitemOpen
  \bibfield  {author} {\bibinfo {author} {\bibfnamefont {P.}~\bibnamefont
  {Calabrese}}, \bibinfo {author} {\bibfnamefont {F.~H.~L.}\ \bibnamefont
  {Essler}}, \ and\ \bibinfo {author} {\bibfnamefont {M.}~\bibnamefont
  {Fagotti}},\ }\href {\doibase 10.1088/1742-5468/2012/07/P07016} {\bibfield
  {journal} {\bibinfo  {journal} {J. Stat. Mech.}\ }\textbf {\bibinfo {volume}
  {2012}},\ \bibinfo {pages} {P07016} (\bibinfo {year}
  {2012}{\natexlab{a}})}\BibitemShut {NoStop}%
\bibitem [{\citenamefont {Calabrese}\ \emph
  {et~al.}(2012{\natexlab{b}})\citenamefont {Calabrese}, \citenamefont
  {Essler},\ and\ \citenamefont {Fagotti}}]{Calabrese2012a}%
  \BibitemOpen
  \bibfield  {author} {\bibinfo {author} {\bibfnamefont {P.}~\bibnamefont
  {Calabrese}}, \bibinfo {author} {\bibfnamefont {F.~H.~L.}\ \bibnamefont
  {Essler}}, \ and\ \bibinfo {author} {\bibfnamefont {M.}~\bibnamefont
  {Fagotti}},\ }\href {\doibase 10.1088/1742-5468/2012/07/P07022} {\bibfield
  {journal} {\bibinfo  {journal} {J. Stat. Mech.}\ }\textbf {\bibinfo {volume}
  {2012}},\ \bibinfo {pages} {P07022} (\bibinfo {year}
  {2012}{\natexlab{b}})}\BibitemShut {NoStop}%
\bibitem [{\citenamefont {Kormos}\ \emph {et~al.}(2014)\citenamefont {Kormos},
  \citenamefont {Collura},\ and\ \citenamefont {Calabrese}}]{Kormos2014}%
  \BibitemOpen
  \bibfield  {author} {\bibinfo {author} {\bibfnamefont {M.}~\bibnamefont
  {Kormos}}, \bibinfo {author} {\bibfnamefont {M.}~\bibnamefont {Collura}}, \
  and\ \bibinfo {author} {\bibfnamefont {P.}~\bibnamefont {Calabrese}},\ }\href
  {\doibase 10.1103/PhysRevA.89.013609} {\bibfield  {journal} {\bibinfo
  {journal} {Phys. Rev. A}\ }\textbf {\bibinfo {volume} {89}},\ \bibinfo
  {pages} {013609} (\bibinfo {year} {2014})}\BibitemShut {NoStop}%
\bibitem [{\citenamefont {Piroli}\ \emph {et~al.}()\citenamefont {Piroli},
  \citenamefont {Calabrese},\ and\ \citenamefont {Essler}}]{Piroli2016}%
  \BibitemOpen
  \bibfield  {author} {\bibinfo {author} {\bibfnamefont {L.}~\bibnamefont
  {Piroli}}, \bibinfo {author} {\bibfnamefont {P.}~\bibnamefont {Calabrese}}, \
  and\ \bibinfo {author} {\bibfnamefont {F.~H.~L.}\ \bibnamefont {Essler}},\
  }\href@noop {} {\ }\Eprint {http://arxiv.org/abs/1604.08141}
  {arXiv:1604.08141} \BibitemShut {NoStop}%
\bibitem [{\citenamefont {Delfino}(2014)}]{Delfino2014}%
  \BibitemOpen
  \bibfield  {author} {\bibinfo {author} {\bibfnamefont {G.}~\bibnamefont
  {Delfino}},\ }\href {http://stacks.iop.org/1751-8121/47/i=40/a=402001}
  {\bibfield  {journal} {\bibinfo  {journal} {J. Phys. A: Math. Theo.}\
  }\textbf {\bibinfo {volume} {47}},\ \bibinfo {pages} {402001} (\bibinfo
  {year} {2014})}\BibitemShut {NoStop}%
\bibitem [{\citenamefont {Delfino}\ and\ \citenamefont {Viti}()}]{Delfino2016}%
  \BibitemOpen
  \bibfield  {author} {\bibinfo {author} {\bibfnamefont {G.}~\bibnamefont
  {Delfino}}\ and\ \bibinfo {author} {\bibfnamefont {J.}~\bibnamefont {Viti}},\
  }\href {http://arxiv.org/abs/1608.07612} {\ }\Eprint
  {http://arxiv.org/abs/1608.07612} {arXiv:1608.07612} \BibitemShut {NoStop}%
\bibitem [{\citenamefont {Berges}\ \emph {et~al.}(2004)\citenamefont {Berges},
  \citenamefont {Bors\'anyi},\ and\ \citenamefont {Wetterich}}]{Berges2004}%
  \BibitemOpen
  \bibfield  {author} {\bibinfo {author} {\bibfnamefont {J.}~\bibnamefont
  {Berges}}, \bibinfo {author} {\bibfnamefont {S.}~\bibnamefont {Bors\'anyi}},
  \ and\ \bibinfo {author} {\bibfnamefont {C.}~\bibnamefont {Wetterich}},\
  }\href {\doibase 10.1103/PhysRevLett.93.142002} {\bibfield  {journal}
  {\bibinfo  {journal} {Phys. Rev. Lett.}\ }\textbf {\bibinfo {volume} {93}},\
  \bibinfo {pages} {142002} (\bibinfo {year} {2004})}\BibitemShut {NoStop}%
\bibitem [{\citenamefont {Fagotti}(2014)}]{Maurizio2014}%
  \BibitemOpen
  \bibfield  {author} {\bibinfo {author} {\bibfnamefont {M.}~\bibnamefont
  {Fagotti}},\ }\href {http://stacks.iop.org/1742-5468/2014/i=3/a=P03016}
  {\bibfield  {journal} {\bibinfo  {journal} {J. Stat. Mech.}\ }\textbf
  {\bibinfo {volume} {2014}},\ \bibinfo {pages} {P03016} (\bibinfo {year}
  {2014})}\BibitemShut {NoStop}%
\bibitem [{\citenamefont {Bertini}\ \emph {et~al.}(2015)\citenamefont
  {Bertini}, \citenamefont {Essler}, \citenamefont {Groha},\ and\ \citenamefont
  {Robinson}}]{Bruno2015}%
  \BibitemOpen
  \bibfield  {author} {\bibinfo {author} {\bibfnamefont {B.}~\bibnamefont
  {Bertini}}, \bibinfo {author} {\bibfnamefont {F.~H.~L.}\ \bibnamefont
  {Essler}}, \bibinfo {author} {\bibfnamefont {S.}~\bibnamefont {Groha}}, \
  and\ \bibinfo {author} {\bibfnamefont {N.~J.}\ \bibnamefont {Robinson}},\
  }\href {\doibase 10.1103/PhysRevLett.115.180601} {\bibfield  {journal}
  {\bibinfo  {journal} {Phys. Rev. Lett.}\ }\textbf {\bibinfo {volume} {115}},\
  \bibinfo {pages} {180601} (\bibinfo {year} {2015})}\BibitemShut {NoStop}%
\bibitem [{\citenamefont {Langen}\ \emph {et~al.}(2016)\citenamefont {Langen},
  \citenamefont {Gasenzer},\ and\ \citenamefont {Schmiedmayer}}]{Tim2016}%
  \BibitemOpen
  \bibfield  {author} {\bibinfo {author} {\bibfnamefont {T.}~\bibnamefont
  {Langen}}, \bibinfo {author} {\bibfnamefont {T.}~\bibnamefont {Gasenzer}}, \
  and\ \bibinfo {author} {\bibfnamefont {J.}~\bibnamefont {Schmiedmayer}},\
  }\href {http://stacks.iop.org/1742-5468/2016/i=6/a=064009} {\bibfield
  {journal} {\bibinfo  {journal} {J. Stat. Mech.}\ }\textbf {\bibinfo {volume}
  {2016}},\ \bibinfo {pages} {064009} (\bibinfo {year} {2016})}\BibitemShut
  {NoStop}%
\bibitem [{\citenamefont {Marcuzzi}\ \emph {et~al.}()\citenamefont {Marcuzzi},
  \citenamefont {Marino}, \citenamefont {Gambassi},\ and\ \citenamefont
  {Silva}}]{Marcuzzi2016}%
  \BibitemOpen
  \bibfield  {author} {\bibinfo {author} {\bibfnamefont {M.}~\bibnamefont
  {Marcuzzi}}, \bibinfo {author} {\bibfnamefont {J.}~\bibnamefont {Marino}},
  \bibinfo {author} {\bibfnamefont {A.}~\bibnamefont {Gambassi}}, \ and\
  \bibinfo {author} {\bibfnamefont {A.}~\bibnamefont {Silva}},\ }\href
  {http://arxiv.org/abs/1609.06361} {\ }\Eprint
  {http://arxiv.org/abs/1609.06361} {arXiv:1609.06361} \BibitemShut {NoStop}%
\bibitem [{\citenamefont {Yin}\ and\ \citenamefont
  {Radzihovsky}(2016)}]{yin_postquench_2016}%
  \BibitemOpen
  \bibfield  {author} {\bibinfo {author} {\bibfnamefont {X.}~\bibnamefont
  {Yin}}\ and\ \bibinfo {author} {\bibfnamefont {L.}~\bibnamefont
  {Radzihovsky}},\ }\href {\doibase 10.1103/PhysRevA.93.033653} {\bibfield
  {journal} {\bibinfo  {journal} {Phys. Rev. A}\ }\textbf {\bibinfo {volume}
  {93}},\ \bibinfo {pages} {033653} (\bibinfo {year} {2016})}\BibitemShut
  {NoStop}%
\bibitem [{\citenamefont {Yin}\ and\ \citenamefont
  {Radzihovsky}()}]{yin_quench_2016}%
  \BibitemOpen
  \bibfield  {author} {\bibinfo {author} {\bibfnamefont {X.}~\bibnamefont
  {Yin}}\ and\ \bibinfo {author} {\bibfnamefont {L.}~\bibnamefont
  {Radzihovsky}},\ }\href {http://arxiv.org/abs/1610.00670} {\bibinfo
  {journal} {arXiv:1610.00670}\ }\BibitemShut {NoStop}%
\bibitem [{\citenamefont {Eisert}\ \emph {et~al.}(2015)\citenamefont {Eisert},
  \citenamefont {Friesdorf},\ and\ \citenamefont {Gogolin}}]{Eisert2015}%
  \BibitemOpen
\bibfield  {journal} {  }\bibfield  {author} {\bibinfo {author} {\bibfnamefont
  {J.}~\bibnamefont {Eisert}}, \bibinfo {author} {\bibfnamefont
  {M.}~\bibnamefont {Friesdorf}}, \ and\ \bibinfo {author} {\bibfnamefont
  {C.}~\bibnamefont {Gogolin}},\ }\href {\doibase 10.1038/nphys3215} {\bibfield
   {journal} {\bibinfo  {journal} {Nat. Phys.}\ }\textbf {\bibinfo {volume}
  {11}},\ \bibinfo {pages} {124} (\bibinfo {year} {2015})}\BibitemShut
  {NoStop}%
\bibitem [{\citenamefont {Essler}\ and\ \citenamefont
  {Fagotti}(2016)}]{Essler2016}%
  \BibitemOpen
  \bibfield  {author} {\bibinfo {author} {\bibfnamefont {F.~H.~L.}\
  \bibnamefont {Essler}}\ and\ \bibinfo {author} {\bibfnamefont
  {M.}~\bibnamefont {Fagotti}},\ }\href
  {http://stacks.iop.org/1742-5468/2016/i=6/a=064002?key=crossref.e9968c9a437b9be91dd2eddbccb6bcee}
  {\bibfield  {journal} {\bibinfo  {journal} {J. Stat. Mech.}\ }\textbf
  {\bibinfo {volume} {2016}},\ \bibinfo {pages} {064002} (\bibinfo {year}
  {2016})}\BibitemShut {NoStop}%
\bibitem [{\citenamefont {Ba{\~{n}}uls}\ \emph {et~al.}(2011)\citenamefont
  {Ba{\~{n}}uls}, \citenamefont {Cirac},\ and\ \citenamefont
  {Hastings}}]{Banuls2011}%
  \BibitemOpen
  \bibfield  {author} {\bibinfo {author} {\bibfnamefont {M.~C.}\ \bibnamefont
  {Ba{\~{n}}uls}}, \bibinfo {author} {\bibfnamefont {J.~I.}\ \bibnamefont
  {Cirac}}, \ and\ \bibinfo {author} {\bibfnamefont {M.~B.}\ \bibnamefont
  {Hastings}},\ }\href {http://link.aps.org/doi/10.1103/PhysRevLett.106.050405}
  {\bibfield  {journal} {\bibinfo  {journal} {Phys. Rev. Lett.}\ }\textbf
  {\bibinfo {volume} {106}},\ \bibinfo {pages} {050405} (\bibinfo {year}
  {2011})}\BibitemShut {NoStop}%
\bibitem [{\citenamefont {Ba{\~{n}}uls}\ \emph {et~al.}(2009)\citenamefont
  {Ba{\~{n}}uls}, \citenamefont {Hastings}, \citenamefont {Verstraete},\ and\
  \citenamefont {Cirac}}]{Banuls2009}%
  \BibitemOpen
  \bibfield  {author} {\bibinfo {author} {\bibfnamefont {M.~C.}\ \bibnamefont
  {Ba{\~{n}}uls}}, \bibinfo {author} {\bibfnamefont {M.~B.}\ \bibnamefont
  {Hastings}}, \bibinfo {author} {\bibfnamefont {F.}~\bibnamefont
  {Verstraete}}, \ and\ \bibinfo {author} {\bibfnamefont {J.~I.}\ \bibnamefont
  {Cirac}},\ }\href {\doibase 10.1103/PhysRevLett.102.240603} {\bibfield
  {journal} {\bibinfo  {journal} {Phys. Rev. Lett.}\ }\textbf {\bibinfo
  {volume} {102}},\ \bibinfo {pages} {240603} (\bibinfo {year}
  {2009})}\BibitemShut {NoStop}%
\bibitem [{\citenamefont {Hastings}\ and\ \citenamefont
  {Mahajan}(2015)}]{Hastings2015}%
  \BibitemOpen
  \bibfield  {author} {\bibinfo {author} {\bibfnamefont {M.~B.}\ \bibnamefont
  {Hastings}}\ and\ \bibinfo {author} {\bibfnamefont {R.}~\bibnamefont
  {Mahajan}},\ }\href {\doibase 10.1103/PhysRevA.91.032306} {\bibfield
  {journal} {\bibinfo  {journal} {Phys. Rev. A}\ }\textbf {\bibinfo {volume}
  {91}},\ \bibinfo {pages} {032306} (\bibinfo {year} {2015})}\BibitemShut
  {NoStop}%
\bibitem [{\citenamefont {Mazza}\ \emph {et~al.}(2014)\citenamefont {Mazza},
  \citenamefont {Collura}, \citenamefont {Kormos},\ and\ \citenamefont
  {Calabrese}}]{Mazza2014}%
  \BibitemOpen
  \bibfield  {author} {\bibinfo {author} {\bibfnamefont {P.~P.}\ \bibnamefont
  {Mazza}}, \bibinfo {author} {\bibfnamefont {M.}~\bibnamefont {Collura}},
  \bibinfo {author} {\bibfnamefont {M.}~\bibnamefont {Kormos}}, \ and\ \bibinfo
  {author} {\bibfnamefont {P.}~\bibnamefont {Calabrese}},\ }\href
  {http://stacks.iop.org/1742-5468/2014/i=11/a=P11016?key=crossref.1a9af174dc010f46ac47e8f3d78dea6d}
  {\bibfield  {journal} {\bibinfo  {journal} {J. Stat. Mech.}\ }\textbf
  {\bibinfo {volume} {2014}},\ \bibinfo {pages} {P11016} (\bibinfo {year}
  {2014})}\BibitemShut {NoStop}%
\bibitem [{\citenamefont {Kormos}\ \emph {et~al.}()\citenamefont {Kormos},
  \citenamefont {Collura}, \citenamefont {Tak{\'{a}}cs},\ and\ \citenamefont
  {Calabrese}}]{Kormos2016}%
  \BibitemOpen
  \bibfield  {author} {\bibinfo {author} {\bibfnamefont {M.}~\bibnamefont
  {Kormos}}, \bibinfo {author} {\bibfnamefont {M.}~\bibnamefont {Collura}},
  \bibinfo {author} {\bibfnamefont {G.}~\bibnamefont {Tak{\'{a}}cs}}, \ and\
  \bibinfo {author} {\bibfnamefont {P.}~\bibnamefont {Calabrese}},\ }\href
  {http://arxiv.org/abs/1604.03571} {\ }\Eprint
  {http://arxiv.org/abs/1604.03571} {arXiv:1604.03571} \BibitemShut {NoStop}%
\bibitem [{\citenamefont {MacDonald}\ \emph {et~al.}(1988)\citenamefont
  {MacDonald}, \citenamefont {Girvin},\ and\ \citenamefont
  {Yoshioka}}]{MacDonald1988}%
  \BibitemOpen
  \bibfield  {author} {\bibinfo {author} {\bibfnamefont {A.~H.}\ \bibnamefont
  {MacDonald}}, \bibinfo {author} {\bibfnamefont {S.~M.}\ \bibnamefont
  {Girvin}}, \ and\ \bibinfo {author} {\bibfnamefont {D.}~\bibnamefont
  {Yoshioka}},\ }\href {http://link.aps.org/doi/10.1103/PhysRevB.37.9753}
  {\bibfield  {journal} {\bibinfo  {journal} {Phys. Rev. B}\ }\textbf {\bibinfo
  {volume} {37}},\ \bibinfo {pages} {9753} (\bibinfo {year}
  {1988})}\BibitemShut {NoStop}%
\bibitem [{\citenamefont {Bravyi}\ \emph {et~al.}(2011)\citenamefont {Bravyi},
  \citenamefont {DiVincenzo},\ and\ \citenamefont {Loss}}]{Bravyi2011}%
  \BibitemOpen
  \bibfield  {author} {\bibinfo {author} {\bibfnamefont {S.}~\bibnamefont
  {Bravyi}}, \bibinfo {author} {\bibfnamefont {D.~P.}\ \bibnamefont
  {DiVincenzo}}, \ and\ \bibinfo {author} {\bibfnamefont {D.}~\bibnamefont
  {Loss}},\ }\href
  {http://linkinghub.elsevier.com/retrieve/pii/S0003491611001059} {\bibfield
  {journal} {\bibinfo  {journal} {Ann. Phys.}\ }\textbf {\bibinfo {volume}
  {326}},\ \bibinfo {pages} {2793} (\bibinfo {year} {2011})}\BibitemShut
  {NoStop}%
\bibitem [{\citenamefont {Grover}\ and\ \citenamefont
  {Fisher}(2014)}]{Grover2014}%
  \BibitemOpen
  \bibfield  {author} {\bibinfo {author} {\bibfnamefont {T.}~\bibnamefont
  {Grover}}\ and\ \bibinfo {author} {\bibfnamefont {M.~P.~A.}\ \bibnamefont
  {Fisher}},\ }\href {\doibase 10.1088/1742-5468/2014/10/P10010} {\bibfield
  {journal} {\bibinfo  {journal} {J. Stat. Mech.}\ }\textbf {\bibinfo {volume}
  {2014}},\ \bibinfo {pages} {P10010} (\bibinfo {year} {2014})}\BibitemShut
  {NoStop}%
\bibitem [{\citenamefont {Garrison}\ \emph {et~al.}()\citenamefont {Garrison},
  \citenamefont {Mishmash},\ and\ \citenamefont {Fisher}}]{Garrison2016}%
  \BibitemOpen
  \bibfield  {author} {\bibinfo {author} {\bibfnamefont {J.~R.}\ \bibnamefont
  {Garrison}}, \bibinfo {author} {\bibfnamefont {R.~V.}\ \bibnamefont
  {Mishmash}}, \ and\ \bibinfo {author} {\bibfnamefont {M.~P.~A.}\ \bibnamefont
  {Fisher}},\ }\href {http://arxiv.org/abs/1606.05650} {\ }\Eprint
  {http://arxiv.org/abs/1606.05650} {arXiv:1606.05650} \BibitemShut {NoStop}%
\bibitem [{\citenamefont {Sachdev}\ and\ \citenamefont
  {Young}(1997)}]{Sachdev1997}%
  \BibitemOpen
  \bibfield  {author} {\bibinfo {author} {\bibfnamefont {S.}~\bibnamefont
  {Sachdev}}\ and\ \bibinfo {author} {\bibfnamefont {A.~P.}\ \bibnamefont
  {Young}},\ }\href {\doibase 10.1103/PhysRevLett.78.2220} {\bibfield
  {journal} {\bibinfo  {journal} {Phys. Rev. Lett.}\ }\textbf {\bibinfo
  {volume} {78}},\ \bibinfo {pages} {2220} (\bibinfo {year}
  {1997})}\BibitemShut {NoStop}%
\bibitem [{\citenamefont {McCoy}\ \emph {et~al.}(1971)\citenamefont {McCoy},
  \citenamefont {Barouch},\ and\ \citenamefont {Abraham}}]{McCoy1971}%
  \BibitemOpen
  \bibfield  {author} {\bibinfo {author} {\bibfnamefont {B.~M.}\ \bibnamefont
  {McCoy}}, \bibinfo {author} {\bibfnamefont {E.}~\bibnamefont {Barouch}}, \
  and\ \bibinfo {author} {\bibfnamefont {D.~B.}\ \bibnamefont {Abraham}},\
  }\href {\doibase 10.1103/PhysRevA.4.2331} {\bibfield  {journal} {\bibinfo
  {journal} {Phys. Rev. A}\ }\textbf {\bibinfo {volume} {4}},\ \bibinfo {pages}
  {2331} (\bibinfo {year} {1971})}\BibitemShut {NoStop}%
\bibitem [{\citenamefont {Hastings}(2004)}]{Hastings2004}%
  \BibitemOpen
  \bibfield  {author} {\bibinfo {author} {\bibfnamefont {M.~B.}\ \bibnamefont
  {Hastings}},\ }\href {\doibase 10.1103/PhysRevLett.93.140402} {\bibfield
  {journal} {\bibinfo  {journal} {Phys. Rev. Lett.}\ }\textbf {\bibinfo
  {volume} {93}},\ \bibinfo {pages} {140402} (\bibinfo {year}
  {2004})}\BibitemShut {NoStop}%
\bibitem [{\citenamefont {McCoy}\ and\ \citenamefont {Wu}()}]{McCoy1973}%
  \BibitemOpen
  \bibfield  {author} {\bibinfo {author} {\bibfnamefont {B.~M.}\ \bibnamefont
  {McCoy}}\ and\ \bibinfo {author} {\bibfnamefont {T.~T.}\ \bibnamefont {Wu}},\
  }\href@noop {} {\emph {\bibinfo {title} {{The Two-Dimensional Ising
  Model}}}}\ (\bibinfo  {publisher} {Harvard University Press Cambridge, MA,
  1973})\BibitemShut {NoStop}%
\bibitem [{\citenamefont {Kitaev}(2001)}]{Kitaev2001}%
  \BibitemOpen
  \bibfield  {author} {\bibinfo {author} {\bibfnamefont {A.~Y.}\ \bibnamefont
  {Kitaev}},\ }\href {\doibase 10.1070/1063-7869/44/10S/S29} {\bibfield
  {journal} {\bibinfo  {journal} {Physics-Uspekhi}\ }\textbf {\bibinfo {volume}
  {44}},\ \bibinfo {pages} {131} (\bibinfo {year} {2001})}\BibitemShut
  {NoStop}%
\bibitem [{\citenamefont {Motrunich}\ \emph {et~al.}(2001)\citenamefont
  {Motrunich}, \citenamefont {Damle},\ and\ \citenamefont
  {Huse}}]{Motrunich2001}%
  \BibitemOpen
  \bibfield  {author} {\bibinfo {author} {\bibfnamefont {O.}~\bibnamefont
  {Motrunich}}, \bibinfo {author} {\bibfnamefont {K.}~\bibnamefont {Damle}}, \
  and\ \bibinfo {author} {\bibfnamefont {D.~A.}\ \bibnamefont {Huse}},\ }\href
  {\doibase 10.1103/PhysRevB.63.224204} {\bibfield  {journal} {\bibinfo
  {journal} {Phys. Rev. B}\ }\textbf {\bibinfo {volume} {63}},\ \bibinfo
  {pages} {224204} (\bibinfo {year} {2001})}\BibitemShut {NoStop}%
\bibitem [{\citenamefont {Read}\ and\ \citenamefont {Green}(2000)}]{Read2000}%
  \BibitemOpen
  \bibfield  {author} {\bibinfo {author} {\bibfnamefont {N.}~\bibnamefont
  {Read}}\ and\ \bibinfo {author} {\bibfnamefont {D.}~\bibnamefont {Green}},\
  }\href {\doibase 10.1103/PhysRevB.61.10267} {\bibfield  {journal} {\bibinfo
  {journal} {Phys. Rev. B}\ }\textbf {\bibinfo {volume} {61}},\ \bibinfo
  {pages} {10267} (\bibinfo {year} {2000})}\BibitemShut {NoStop}%
\bibitem [{\citenamefont {{Motrunich}}(2012)}]{Motrunich2012}%
  \BibitemOpen
  \bibfield  {author} {\bibinfo {author} {\bibfnamefont {O.}~\bibnamefont
  {{Motrunich}}},\ }in\ \href
  {http://meetings.aps.org/link/BAPS.2012.MAR.P23.13} {\emph {\bibinfo
  {booktitle} {\rm {Proceedings of APS March Meeting}}}}\ (\bibinfo {year}
  {2012})\BibitemShut {NoStop}%
\end{thebibliography}
%

\end{document}